\documentclass[twocolumn,letterpaper,aps,prd,longbibliography,superscriptaddress,nofootinbib,floatfix]{revtex4-1}

\usepackage{graphicx}	
\usepackage{amsmath}
\usepackage{xspace}	

\newcommand{\pt}{\mbox{$p_T$}\xspace}

\newcommand{\sqrts}{\mbox{$\sqrt{s}$}\xspace}
\newcommand{\jpsi}{\mbox{$J/\psi$}\xspace}

\newcommand{\pp}{\mbox{$p$+$p$}\xspace}

\begin{document}

\title{Cross section and transverse single-spin asymmetry of muons from 
open heavy-flavor decays in polarized $p$+$p$ collisions at $\sqrt{s}=200$~GeV}


\newcommand{\abilene}{Abilene Christian University, Abilene, Texas 79699, USA}
\newcommand{\augie}{Department of Physics, Augustana University, Sioux Falls, South Dakota 57197, USA}
\newcommand{\banaras}{Department of Physics, Banaras Hindu University, Varanasi 221005, India}
\newcommand{\barc}{Bhabha Atomic Research Centre, Bombay 400 085, India}
\newcommand{\baruch}{Baruch College, City University of New York, New York, New York, 10010 USA}
\newcommand{\bnlcoll}{Collider-Accelerator Department, Brookhaven National Laboratory, Upton, New York 11973-5000, USA}
\newcommand{\bnlphys}{Physics Department, Brookhaven National Laboratory, Upton, New York 11973-5000, USA}
\newcommand{\caucr}{University of California-Riverside, Riverside, California 92521, USA}
\newcommand{\charlesczech}{Charles University, Ovocn\'{y} trh 5, Praha 1, 116 36, Prague, Czech Republic}
\newcommand{\chonbuk}{Chonbuk National University, Jeonju, 561-756, Korea}
\newcommand{\ciae}{Science and Technology on Nuclear Data Laboratory, China Institute of Atomic Energy, Beijing 102413, People's Republic of China}
\newcommand{\cns}{Center for Nuclear Study, Graduate School of Science, University of Tokyo, 7-3-1 Hongo, Bunkyo, Tokyo 113-0033, Japan}
\newcommand{\colorado}{University of Colorado, Boulder, Colorado 80309, USA}
\newcommand{\columbia}{Columbia University, New York, New York 10027 and Nevis Laboratories, Irvington, New York 10533, USA}
\newcommand{\czechtech}{Czech Technical University, Zikova 4, 166 36 Prague 6, Czech Republic}
\newcommand{\debrecen}{Debrecen University, H-4010 Debrecen, Egyetem t{\'e}r 1, Hungary}
\newcommand{\elte}{ELTE, E{\"o}tv{\"o}s Lor{\'a}nd University, H-1117 Budapest, P{\'a}zm{\'a}ny P.~s.~1/A, Hungary}
\newcommand{\eszterhazy}{Eszterh\'azy K\'aroly University, K\'aroly R\'obert Campus, H-3200 Gy\"ngy\"os, M\'atrai \'ut 36, Hungary}
\newcommand{\ewha}{Ewha Womans University, Seoul 120-750, Korea}
\newcommand{\fsu}{Florida State University, Tallahassee, Florida 32306, USA}
\newcommand{\gsu}{Georgia State University, Atlanta, Georgia 30303, USA}
\newcommand{\hanyang}{Hanyang University, Seoul 133-792, Korea}
\newcommand{\hiroshima}{Hiroshima University, Kagamiyama, Higashi-Hiroshima 739-8526, Japan}
\newcommand{\howard}{Department of Physics and Astronomy, Howard University, Washington, DC 20059, USA}
\newcommand{\ihepprot}{IHEP Protvino, State Research Center of Russian Federation, Institute for High Energy Physics, Protvino, 142281, Russia}
\newcommand{\illuiuc}{University of Illinois at Urbana-Champaign, Urbana, Illinois 61801, USA}
\newcommand{\inrras}{Institute for Nuclear Research of the Russian Academy of Sciences, prospekt 60-letiya Oktyabrya 7a, Moscow 117312, Russia}
\newcommand{\instpasczech}{Institute of Physics, Academy of Sciences of the Czech Republic, Na Slovance 2, 182 21 Prague 8, Czech Republic}
\newcommand{\isu}{Iowa State University, Ames, Iowa 50011, USA}
\newcommand{\jaea}{Advanced Science Research Center, Japan Atomic Energy Agency, 2-4 Shirakata Shirane, Tokai-mura, Naka-gun, Ibaraki-ken 319-1195, Japan}
\newcommand{\jyvaskyla}{Helsinki Institute of Physics and University of Jyv{\"a}skyl{\"a}, P.O.Box 35, FI-40014 Jyv{\"a}skyl{\"a}, Finland}
\newcommand{\kek}{KEK, High Energy Accelerator Research Organization, Tsukuba, Ibaraki 305-0801, Japan}
\newcommand{\korea}{Korea University, Seoul, 136-701, Korea}
\newcommand{\kurchatov}{National Research Center ``Kurchatov Institute", Moscow, 123098 Russia}
\newcommand{\kyoto}{Kyoto University, Kyoto 606-8502, Japan}
\newcommand{\labllr}{Laboratoire Leprince-Ringuet, Ecole Polytechnique, CNRS-IN2P3, Route de Saclay, F-91128, Palaiseau, France}
\newcommand{\lahorelums}{Physics Department, Lahore University of Management Sciences, Lahore 54792, Pakistan}
\newcommand{\lawllnl}{Lawrence Livermore National Laboratory, Livermore, California 94550, USA}
\newcommand{\losalamos}{Los Alamos National Laboratory, Los Alamos, New Mexico 87545, USA}
\newcommand{\lund}{Department of Physics, Lund University, Box 118, SE-221 00 Lund, Sweden}
\newcommand{\maryland}{University of Maryland, College Park, Maryland 20742, USA}
\newcommand{\mass}{Department of Physics, University of Massachusetts, Amherst, Massachusetts 01003-9337, USA}
\newcommand{\michigan}{Department of Physics, University of Michigan, Ann Arbor, Michigan 48109-1040, USA}
\newcommand{\muhlenberg}{Muhlenberg College, Allentown, Pennsylvania 18104-5586, USA}
\newcommand{\myongji}{Myongji University, Yongin, Kyonggido 449-728, Korea}
\newcommand{\nagasaki}{Nagasaki Institute of Applied Science, Nagasaki-shi, Nagasaki 851-0193, Japan}
\newcommand{\nara}{Nara Women's University, Kita-uoya Nishi-machi Nara 630-8506, Japan}
\newcommand{\natmephi}{National Research Nuclear University, MEPhI, Moscow Engineering Physics Institute, Moscow, 115409, Russia}
\newcommand{\newmex}{University of New Mexico, Albuquerque, New Mexico 87131, USA}
\newcommand{\nmsu}{New Mexico State University, Las Cruces, New Mexico 88003, USA}
\newcommand{\ohio}{Department of Physics and Astronomy, Ohio University, Athens, Ohio 45701, USA}
\newcommand{\ornl}{Oak Ridge National Laboratory, Oak Ridge, Tennessee 37831, USA}
\newcommand{\orsay}{IPN-Orsay, Univ.~Paris-Sud, CNRS/IN2P3, Universit\'e Paris-Saclay, BP1, F-91406, Orsay, France}
\newcommand{\peking}{Peking University, Beijing 100871, People's Republic of China}
\newcommand{\pnpi}{PNPI, Petersburg Nuclear Physics Institute, Gatchina, Leningrad region, 188300, Russia}
\newcommand{\riken}{RIKEN Nishina Center for Accelerator-Based Science, Wako, Saitama 351-0198, Japan}
\newcommand{\rikjrbrc}{RIKEN BNL Research Center, Brookhaven National Laboratory, Upton, New York 11973-5000, USA}
\newcommand{\rikkyo}{Physics Department, Rikkyo University, 3-34-1 Nishi-Ikebukuro, Toshima, Tokyo 171-8501, Japan}
\newcommand{\saispbstu}{Saint Petersburg State Polytechnic University, St.~Petersburg, 195251 Russia}
\newcommand{\seoulnat}{Department of Physics and Astronomy, Seoul National University, Seoul 151-742, Korea}
\newcommand{\stonybrkc}{Chemistry Department, Stony Brook University, SUNY, Stony Brook, New York 11794-3400, USA}
\newcommand{\stonycrkp}{Department of Physics and Astronomy, Stony Brook University, SUNY, Stony Brook, New York 11794-3800, USA}
\newcommand{\sungskku}{Sungkyunkwan University, Suwon, 440-746, Korea}
\newcommand{\tenn}{University of Tennessee, Knoxville, Tennessee 37996, USA}
\newcommand{\titech}{Department of Physics, Tokyo Institute of Technology, Oh-okayama, Meguro, Tokyo 152-8551, Japan}
\newcommand{\tsukuba}{Center for Integrated Research in Fundamental Science and Engineering, University of Tsukuba, Tsukuba, Ibaraki 305, Japan}
\newcommand{\vandy}{Vanderbilt University, Nashville, Tennessee 37235, USA}
\newcommand{\weizmann}{Weizmann Institute, Rehovot 76100, Israel}
\newcommand{\wigner}{Institute for Particle and Nuclear Physics, Wigner Research Centre for Physics, Hungarian Academy of Sciences (Wigner RCP, RMKI) H-1525 Budapest 114, POBox 49, Budapest, Hungary}
\newcommand{\yonsei}{Yonsei University, IPAP, Seoul 120-749, Korea}
\newcommand{\zagreb}{Department of Physics, Faculty of Science, University of Zagreb, Bijeni\v{c}ka c.~32 HR-10002 Zagreb, Croatia}
\affiliation{\abilene}
\affiliation{\augie}
\affiliation{\banaras}
\affiliation{\barc}
\affiliation{\baruch}
\affiliation{\bnlcoll}
\affiliation{\bnlphys}
\affiliation{\caucr}
\affiliation{\charlesczech}
\affiliation{\chonbuk}
\affiliation{\ciae}
\affiliation{\cns}
\affiliation{\colorado}
\affiliation{\columbia}
\affiliation{\czechtech}
\affiliation{\debrecen}
\affiliation{\elte}
\affiliation{\eszterhazy}
\affiliation{\ewha}
\affiliation{\fsu}
\affiliation{\gsu}
\affiliation{\hanyang}
\affiliation{\hiroshima}
\affiliation{\howard}
\affiliation{\ihepprot}
\affiliation{\illuiuc}
\affiliation{\inrras}
\affiliation{\instpasczech}
\affiliation{\isu}
\affiliation{\jaea}
\affiliation{\jyvaskyla}
\affiliation{\kek}
\affiliation{\korea}
\affiliation{\kurchatov}
\affiliation{\kyoto}
\affiliation{\labllr}
\affiliation{\lahorelums}
\affiliation{\lawllnl}
\affiliation{\losalamos}
\affiliation{\lund}
\affiliation{\maryland}
\affiliation{\mass}
\affiliation{\michigan}
\affiliation{\muhlenberg}
\affiliation{\myongji}
\affiliation{\nagasaki}
\affiliation{\nara}
\affiliation{\natmephi}
\affiliation{\newmex}
\affiliation{\nmsu}
\affiliation{\ohio}
\affiliation{\ornl}
\affiliation{\orsay}
\affiliation{\peking}
\affiliation{\pnpi}
\affiliation{\riken}
\affiliation{\rikjrbrc}
\affiliation{\rikkyo}
\affiliation{\saispbstu}
\affiliation{\seoulnat}
\affiliation{\stonybrkc}
\affiliation{\stonycrkp}
\affiliation{\sungskku}
\affiliation{\tenn}
\affiliation{\titech}
\affiliation{\tsukuba}
\affiliation{\vandy}
\affiliation{\weizmann}
\affiliation{\wigner}
\affiliation{\yonsei}
\affiliation{\zagreb}
\author{C.~Aidala} \affiliation{\losalamos} \affiliation{\michigan} 
\author{N.N.~Ajitanand} \affiliation{\stonybrkc} 
\author{Y.~Akiba} \email[PHENIX Spokesperson: ]{akiba@rcf.rhic.bnl.gov} \affiliation{\riken} \affiliation{\rikjrbrc} 
\author{R.~Akimoto} \affiliation{\cns} 
\author{J.~Alexander} \affiliation{\stonybrkc} 
\author{M.~Alfred} \affiliation{\howard} 
\author{K.~Aoki} \affiliation{\kek} \affiliation{\riken} 
\author{N.~Apadula} \affiliation{\isu} \affiliation{\stonycrkp} 
\author{H.~Asano} \affiliation{\kyoto} \affiliation{\riken} 
\author{E.T.~Atomssa} \affiliation{\stonycrkp} 
% ...
\collaboration{PHENIX Collaboration} \noaffiliation

\date{\today}


\begin{abstract}

The cross section and transverse single-spin asymmetries of $\mu^{-}$ 
and $\mu^{+}$ from open heavy-flavor decays in polarized $p$+$p$ 
collisions at $\sqrt{s}=200$~GeV were measured by the PHENIX experiment 
during 2012 at the Relativistic Heavy Ion Collider.  Because 
heavy-flavor production is dominated by gluon-gluon interactions at 
$\sqrt{s}=200$~GeV, these measurements offer a unique opportunity to 
obtain information on the trigluon correlation functions.  The 
measurements are performed at forward and backward rapidity 
($1.4<|y|<2.0$) over the transverse momentum range of $1.25<p_T<7$ 
GeV/$c$ for the cross section and $1.25<p_T<5$ GeV/$c$ for the asymmetry 
measurements. The obtained cross section is compared to a 
fixed-order-plus-next-to-leading-log perturbative-quantum-chromodynamics 
calculation. The asymmetry results are consistent with zero within 
uncertainties, and a model calculation based on twist-3 three-gluon 
correlations agrees with the data.

\end{abstract}

	
\maketitle


\section{Introduction}

Transverse single-spin asymmetry (TSSA) phenomena have gained 
substantial attention in both experimental and theoretical studies in 
recent years. The existence of TSSAs has been well established in the 
production of light mesons at forward rapidity in transversely polarized 
$p$$+$$p$ collisions at energies ranging from the Zero Gradient 
Synchrotron up to the Relativistic Heavy Ion Collider (RHIC). 
Surprisingly large but oppositely-signed TSSA results were first 
observed in $\pi^{+}$ and $\pi^{-}$ production at large Feynman-$x$ 
($x_F$) in transversely polarized \pp collisions at $\sqrt{s} = 4.9$ 
GeV~\cite{klem:1976ui}. These results surprised the 
quantum-chromodynamics (QCD) community because they disagreed with the 
expectation from the naive perturbative QCD of very small spin 
asymmetries~\cite{Kane:1978nd}. The large TSSA of pion production has 
been subsequently observed in hadronic collisions over a range of 
energies extending up to $\sqrt{s} =$ 500 GeV for $\pi^0$ ($\sqrt{s} =$ 
200 GeV for 
$\pi^{\pm}$)~\cite{Allgower:2002qi,Antille:1980th,Adams:1991rw,Adams:1991cs,Arsene:2008mi,Adams:2003fx,Abelev:2008qb,Mondal:2014vla,Heppelmann:2016siw,Adare:2013ekj}. 
Furthermore, TSSA in $\eta$ meson production has also been studied at 
forward rapidity~\cite{Adare:2014qzo,Adamczyk:2012xd}. The results are 
consistent with the observed $\pi^0$ asymmetries at various energies in 
the overlapping $x_F$ regions. Two theoretical formalisms within the 
perturbative QCD framework have been proposed to explain the origin of 
these large TSSAs at forward rapidity. Both formalisms connect the TSSA 
to the transverse motion of the partons inside the 
transversely-polarized nucleon and/or to spin-dependent quark 
fragmentation.

One framework is based on the transverse-momentum-dependent (TMD) parton 
distribution and fragmentation functions, called TMD factorization. The 
initial state contributions are originating from the Sivers 
function~\cite{Sivers:1989cc,Sivers:1990fh}, which describes the 
correlation between the transverse spin of the nucleon and the parton 
transverse momentum in the initial state. The final state contribution 
originates from the quark transversity distribution and the Collins 
\cite{Collins:1992kk} fragmentation function, which describes the 
fragmentation of a transversely polarized quark into a final state 
hadron with nonzero transverse momentum relative to the parton 
direction. This framework requires two observed scales where only one 
needs to be hard and both effects have been observed in SIDIS 
measurements~\cite{Airapetian:2010ds,Adolph:2012sn}. However, TMD 
factorization cannot be used in the interpretation of hadron production 
in $p$$+$$p$ collisions as only one hard scale is available 
\cite{Rogers:2010dm}.

A second framework, applicable to our study, follows the QCD collinear 
factorization approach. The collinear, higher-twist effects become more 
important in generating a large TSSA when there is only one observed 
momentum scale that is much larger than the nonperturbative hadronic 
scale $\Lambda_{QCD}\approx 200$ MeV~\cite{Efremov:1984ip,Qiu:1991pp}. A 
large TSSA can be generated from the twist-3, transverse-spin-dependent, 
multi-parton correlation functions in the initial state or fragmentation 
functions in the final state.

At RHIC energies, gluon-gluon interaction processes dominate heavy quark 
production~\cite{Norrbin:2000zc}, so heavy quarks serve to isolate the 
gluon contribution to the asymmetries. PHENIX has measured the TSSA 
($A_N$) of $J/\psi$ in central and forward rapidity~\cite{Adare:2010bd}. 
Theoretical predictions of the $J/\psi$ single-spin asymmetry are 
complicated by the lack of good understanding of $J/\psi$ production 
mechanism~\cite{Yuan:2008vn}. In addition, there are feed-down 
contributions from higher resonance states in inclusive $J/\psi$ 
production~\cite{Adare:2011vq}. On the other hand, the effect of pure 
gluonic correlation functions on $D$-meson production in transversely 
polarized \pp collisions has been extensively studied within the twist-3 
mechanism in the framework of collinear 
factorization~\cite{Koike:2011mb,Kang:2008ih}. However, it is difficult 
to constrain the trigluon correlation functions due to the lack of 
experimental results. Future measurements including $D$-meson production 
are proposed at the Large Hadron Collider~\cite{Brodsky:2012vg}.

This paper reports on measurements of the cross section and TSSA for 
muons from open heavy-flavor decays in polarized \pp collisions at 
$\sqrt{s}=200~{\rm GeV}$. Results are presented for muons from 
semi-leptonic decays of open heavy-flavor hadrons, mainly 
$D\rightarrow \mu + X$ and $B\rightarrow \mu + X$, in the forward and backward 
rapidity regions ($1.4<|y|<2.0$); the accessible momentum fraction of 
gluons in the proton is 0.0125--0.0135 and 0.08--0.14 in the backward 
($x_F<0$) and forward ($x_F>0$) regions with respect to the polarized 
beam direction, respectively. Sec.~\ref{sec:phenix} describes the RHIC 
polarized proton beams and the PHENIX experimental setup. The detailed 
analysis of muons from open heavy-flavor, including cross sections and 
TSSAs, will be described in Sec.~\ref{sec:analysis} and the results will 
be presented in Sec.~\ref{sec:results}. Finally, a discussion of the 
results and their possible implications will be provided in 
Sec.~\ref{sec:discussion}.

\section{Experimental Setup}
\label{sec:phenix}

\subsection{The PHENIX experiment}

\begin{figure}[thb]
\includegraphics[width=0.98\linewidth]{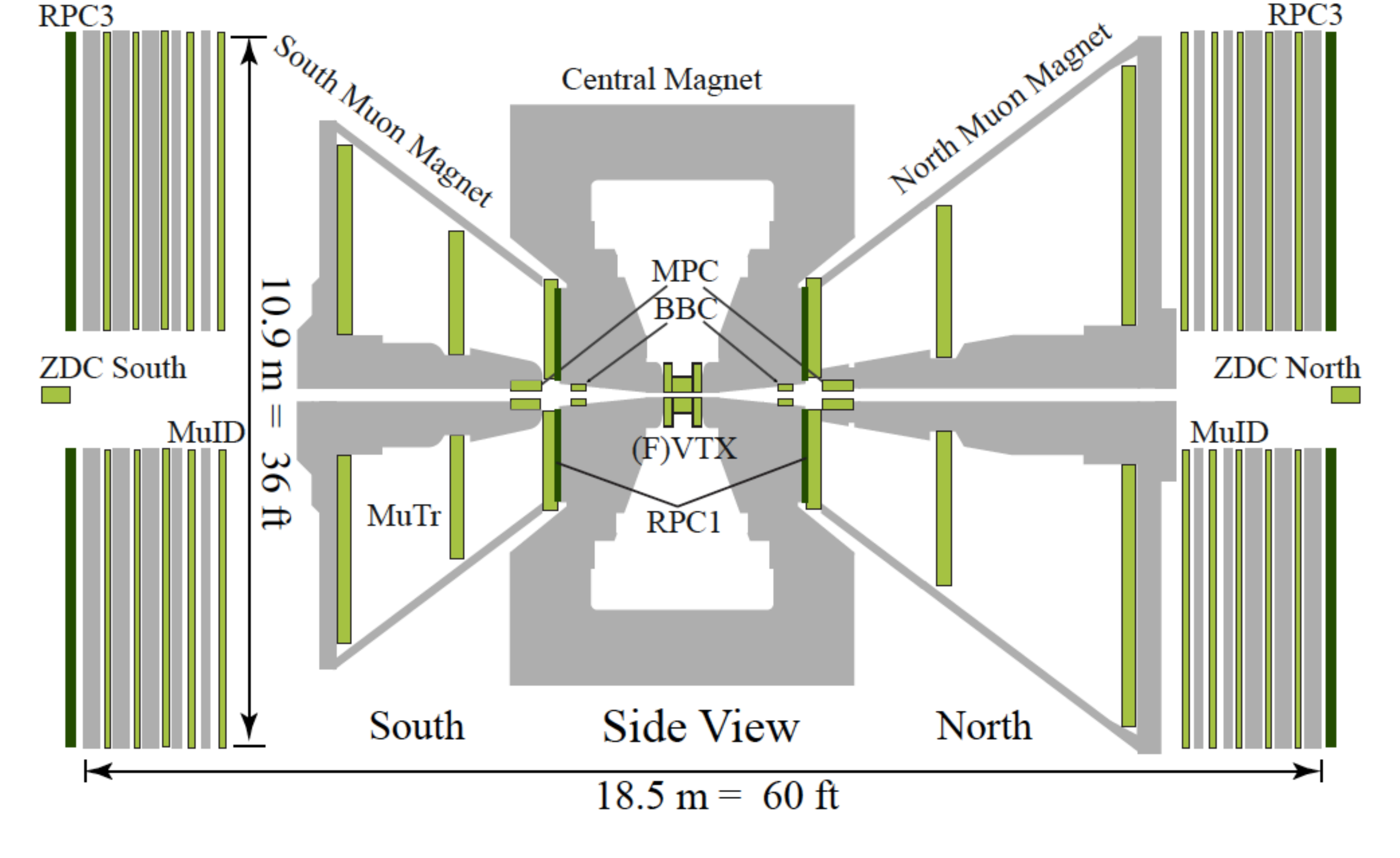}
\caption{\label{fig:PHENIX}
Side view of the PHENIX detector in the 2012 run}
\end{figure}

The PHENIX detector comprises two central arms at midrapidity and 
two muon arms at forward and backward rapidity~\cite{Adcox:2003zm}. As 
shown in Fig.~\ref{fig:PHENIX}, two muon spectrometers cover the full 
azimuthal angle in the pseudorapidity range $1.2<\eta<2.4$ (north arm) 
and $-2.2<\eta<-1.2$ (south arm). In front of each muon arm, there is 
about 7 interaction lengths ($\lambda_{I}$) of copper-and-iron absorber 
which provides a rejection factor of 1000 for charged pions, and an 
additional stainless-steel absorber (2 $\lambda_{I}$ in total) installed 
in 2011 contributes to further suppress hadronic 
background~\cite{Akikawa:2003zs,Adachi:2013qha}.  Each muon arm has 
three stations of cathode strip chambers, muon tracker (MuTr), for 
momentum measurement and five layers (labeled from Gap0 to Gap4) of 
proportional tube planes, muon identifier (MuID), for muon 
identification.  Each MuID gap comprises a plane of absorber 
($\sim1\lambda_{I}$) and two planes of Iarrocci tubes whose orientation 
is along either the horizontal or the vertical direction in each plane. 
The MuID also provides a trigger for events containing one or more muon 
candidates.

The minimum bias (MB) trigger is provided by the beam-beam counters 
(BBC)~\cite{Allen:2003zt}, which comprise two arrays of 64 quartz 
\v{C}erenkov detectors to detect charged particles at high pseudorapidity. 
Each detector is located at $z=\pm144~{\rm cm}$ from the interaction 
point, and covers the pseudorapidity range $3.1<|\eta|<3.9$. The BBC 
also determines the collision-vertex position ($z_{\rm vtx}$) along the 
beam axis, with a resolution of roughly 2~cm in \pp collisions.

\subsection{RHIC polarized beams}

RHIC is a unique, polarized \pp collider located at Brookhaven 
National Laboratory.  RHIC comprises two counter-circulating storage 
rings, in each of which as many as 120 polarized-proton bunches can be 
accelerated to a maximum energy of 255 GeV per proton.

In the 2012 run, the beam injected into RHIC typically consisted of 109 
filled bunches in each ring. The bunches collided with a one-to-one 
correspondence with a 106 ns separation.  Pre-defined polarization 
patterns for every 8 bunches were changed fill-by-fill in order to 
reduce systematic effects. Two polarimeters are used to determine the 
beam polarizations. One is a hydrogen-jet polarimeter, which takes 
several hours to measure the absolute polarization~\cite{Okada:2005gu}. 
The other is a fast, proton-carbon polarimeter which measures relative 
changes in the magnitude of the polarization and any variations across 
the transverse profile of the beam several times per 
fill~\cite{Nakagawa:2008zzb,Huang:2006cs}. During the $\sqrt{s}=200$~GeV 
run in 2012, the polarization direction in the PHENIX interaction region 
was transverse. The average clockwise-beam (known as blue beam) 
polarization for the data used in this analysis was $P=0.64{\pm}0.03$, 
and the average counter-clockwise-beam (yellow beam) polarization was 
$P=0.59{\pm}0.03$. There is a 3.4\% global scale uncertainty in the 
measured $A_N$ due to the polarization uncertainty.

\section{Data Analysis}
\label{sec:analysis}

\subsection{Data set}

We analyzed a data set from transversely polarized \pp collisions at 
$\sqrts=200~{\rm GeV}$ collected with the PHENIX detector in 2012 with 
an integrated luminosity of 9.2~pb$^{-1}$. These data have been recorded 
by using the MuID trigger in coincidence with the BBC trigger. The BBC 
trigger requires at least one hit in both BBCs. The BBC trigger 
efficiency for MB \pp events (events containing muons from open 
heavy-flavor) is 55\% (79\%)~\cite{Adler:2003pb} with the van der Meer 
scan technique~\cite{vanderMeerscan}. The MuID trigger serves to select 
events containing at least one MuID track reaching Gap3 or Gap4.

\subsection{Yield of muons from open heavy-flavor}

PHENIX has reported several measurements of muons from open heavy-flavor 
decays in various collision systems~\cite{Adare:2012px,Adare:2013lkk}. 
Similar methods developed in the previous analyses for background 
estimation are used in this analysis. Due to the benefit of the 
additional absorber material, the measurement of positively-charged 
muons from open heavy-flavor decays is possible in PHENIX for the first 
time with these data.

\subsubsection{Muon-candidate selection}

We choose tracks penetrating through all the MuID gaps as good muon 
candidates from events for which the BBC $z$-vertex is within 
$\pm25~{\rm cm}$. Track quality cuts, shown in Table~\ref{tab:cuts}, are 
also required to reject background tracks. DG0 is the distance between 
the projected positions of a MuTr track and a MuID track at the $z$ 
position of the MuID Gap0. DDG0 is the angular difference between the 
two projected positions used in the DG0. $r_{\rm ref}$ is the distance 
between the interaction point and a projected position of a MuID track 
at $z=0$. $p\cdot(\theta_{\rm MuTr} - \theta_{\rm vtx})$ is the polar 
scattering angle of a track inside the absorber scaled by the momentum, 
where $\theta_{\rm vtx}$ is the angle at the vertex and 
$\theta_{\rm MuTr}$ is the angle at the MuTr Station 1. Two cuts, on 
$p\cdot(\theta_{\rm MuTr} - \theta_{\rm vtx})$ and $\chi^{2}$ at 
$z_{\rm vtx}$, are effective for rejecting tracks suffering from large multiple 
scattering or decaying to muons inside the absorber. Track quality cuts 
are determined with the help of a Monte Carlo simulation with {\sc 
geant4}~\cite{Agostinelli:2002hh}; the cut values vary with the momentum 
of the track.

\begin{table}[tbh]
\caption{\label{tab:cuts}
Track selection cuts used in this analysis. Cut values vary with the \pt 
of track; those shown here are for the lowest-\pt bin 
($1.25<p_{T}<1.5~{\rm GeV}/c$).}
\begin{ruledtabular} \begin{tabular}{c}
DG0 $<20~{\rm cm}$ (South), 10~{\rm cm} (North)\\
DDG0 $<8~{\rm deg.}$\\
$r_{\rm ref}<125~{\rm cm}$\\
number of hits in MuTr $>12$, $\chi^{2}_{\rm MuTr}/ndf < 10$ \\
number of hits in MuID $>6$, $\chi^{2}_{\rm MuID}/ndf < 5$ \\
$p\cdot(\theta_{\rm MuTr} - \theta_{\rm vtx})<0.2~{\rm rad}\cdot{\rm GeV}/c$\\
$\chi^{2}$ of track projection to $z_{\rm vtx}<4$
\end{tabular} \end{ruledtabular}
\end{table}

In this analysis, we also use tracks that stopped at MuID Gap3 for 
background estimation, although these tracks are not considered as muon 
candidates. After applying a proper $p_{z}$ cut ($p_{z}\sim3.8~{\rm 
GeV}/c$), we obtain a data sample enriched in hadrons (called stopped 
hadrons)~\cite{Adare:2012px}. These tracks are used to determine the 
punch-through hadron background which arises from hadrons traversing 
through all MuID layers without decay; this background is described in 
more detail in the next section.

\subsubsection{Background estimation}
\label{sec:bkg_estimation}

The primary sources of background tracks are charged pions and kaons. 
Decay muons from $\pi^{\pm}$ and $K^{\pm}$ are the dominant background 
for $\pt<5~{\rm GeV}/c$, while the fraction of punch-through hadrons 
becomes larger at $\pt>5~{\rm GeV}/c$. Another background component is 
muons from \jpsi decays. The contribution from \jpsi decay is small in 
the low-\pt region but increases up to 20\% of muons from inclusive 
heavy-flavor decays at $\pt\sim5~{\rm GeV}/c$.  Backgrounds from light 
resonances ($\phi$, $\rho$, and $\omega$) or other quarkonium states 
($\chi_{c}$, $\psi^\prime$, and $\Upsilon$) are 
negligible~\cite{Adare:2012px,Adare:2010de}. Therefore, the number of 
muons from open heavy-flavor decays is obtained as,

\begin{equation}
N_{\rm HF} = N_{\rm incl}/\varepsilon_{\rm trig} - N_{\rm DM} - N_{\rm PH} - N_{J/\psi\to\mu},
\end{equation}
where $N_{\rm HF}$ is the number of muons from open heavy-flavor decays, 
$N_{\rm incl}$ is the number of muon candidates passing through all 
track quality cuts in Table~\ref{tab:cuts}, $\varepsilon_{\rm trig}$ is 
the trigger efficiency of the MuID trigger, $N_{\rm DM}$ is the 
estimated number of decay muons from $\pi^{\pm}$ and $K^{\pm}$, $N_{\rm PH}$ 
is the estimated number of punch-through hadrons, and 
$N_{J/\psi\to\mu}$ is the estimated number of muons from \jpsi decay. 
The trigger efficiency correction should be taken into account before 
subtracting the background, because the simulation of the backgrounds 
does not include any inefficiency of the MuID trigger. The MuID trigger 
efficiency is evaluated with data by measuring the fraction of MUID 
triggers in non-MUID triggered events containing tracks at MuID Gap3 or 
Gap4.

To estimate the hadronic background ($N_{\rm DM}$ and $N_{\rm PH}$), 
the hadron-cocktail method, developed for the previous 
analysis~\cite{Adare:2012px,Adare:2010de}, is used. Initial particle 
distributions for the hadron-cocktail simulation are estimated from 
measurements of charged pions and kaons at 
midrapidity~\cite{Adare:2011vy,Agakishiev:2011dc}. The {\sc pythia} 
event generator~\cite{Sjostrand:2006za} is used to extrapolate the \pt 
spectra at midrapidity to the forward rapidity region. To 
obtain enough statistics of reconstructed tracks in the high-\pt region, 
a $p_T^{3}$ weight is applied to the estimated \pt spectra for the 
simulation and the simulation output is reweighted by $1/p_{T}^{3}$ for 
a proper comparison with the data. Based on these initial hadron 
distributions, a full chain of detector simulation with 
{\sc geant4}~\cite{Agostinelli:2002hh} and track reconstruction is performed. 
Due to uncertainties in the estimation of input distributions and 
hadron-shower simulation with the thick absorber in front of the MuTr, 
an additional, data-driven, tuning procedure of the simulation is needed 
to determine the background more precisely. Two methods, described 
below, are used to tune the hadron-cocktail simulation:

\begin{description}

\item[Normalized $z_{\rm vtx}$ distribution:] The $z_{\rm vtx}$ 
distribution of tracks ($dN_{\mu}/dz_{\rm vtx}$) normalized by the 
$z_{\rm vtx}$ distribution of MB events ($dN_{\rm evt}/dz_{\rm vtx}$) 
provides a good constraint on the decay muon background.  Because the 
distance from $z_{\rm vtx}$ to the front absorber is relatively short 
compared to the decay length of $\pi^{\pm}$ and $K^{\pm}$, the 
production of decay muons shows a linear dependence on $z_{\rm vtx}$. 
Therefore, the number of decay muons can be estimated by matching the 
slope in the normalized $z_{\rm vtx}$ distribution at MuID Gap4 for each 
\pt bin. More details are described in~\cite{Adare:2012px}.

\item[Stopped hadrons:] Hadrons stopping at MuID Gap3 can be removed 
with an appropriate momentum cut ($p_{z}\sim3.8~{\rm GeV}/c$) as 
described in the previous section. The remaining stopped muons are less 
than 10\% in the tracks at MuID Gap3, based on the simulation study. The 
punch-through hadron background at the last MuID gap can be estimated by 
matching the \pt distribution of stopped hadrons at MuID Gap3.

\end{description}

After tuning the hadron-cocktail simulation, the decay muons ($N_{\rm DM}$) 
from the normalized $z_{\rm vtx}$ distribution matching and the 
punch-through hadrons ($N_{\rm PH}$) from the stopped-hadron matching 
are combined for the final estimate of the background from light 
hadrons. For the decay muons at $\pt>3~{\rm GeV}/c$ and the 
punch-through hadrons, the difference between the two methods of tuning 
is assigned as the systematic uncertainty. More details on the 
hadron-cocktail simulation and the tuning procedure are given 
in~\cite{Adare:2012px}.

\begin{figure}[thb]
\includegraphics[width=1.0\linewidth]{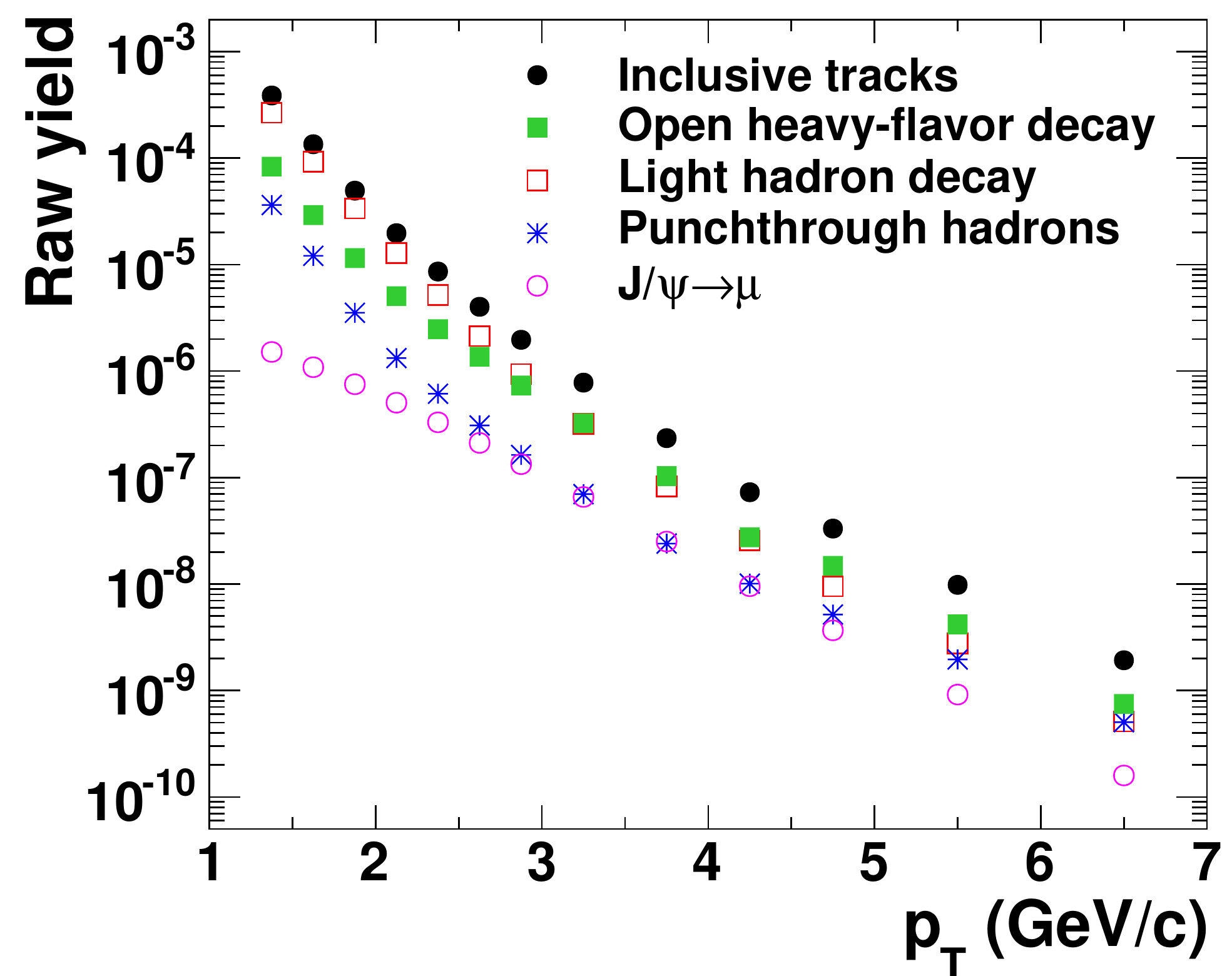}
\caption{\label{fig:gap4_comp} 
\pt spectra of inclusive muon candidates and background sources from the 
hadron-cocktail simulation after \pt-dependent tuning.}
\end{figure}

Muons from \jpsi decays are also subtracted in order to obtain the 
number of muons from open heavy-flavor decays. From the measurement of 
the \jpsi invariant cross section in the forward 
region~\cite{Adare:2011vq} and a decay simulation, the number of muons 
from \jpsi decay ($N_{J/\psi\to\mu}$) can be 
estimated~\cite{Adare:2010de}. The contribution of muons from \jpsi to 
the muons from inclusive heavy-flavor decays is $\sim2\%$ at low \pt and 
increases up to $\sim20\%$ at $\pt>5~{\rm GeV}/c$.  Because there is a 
$B\to\jpsi$ contribution in the inclusive \jpsi measurement, a fraction 
of $B$ is included in $N_{J/\psi\to\mu}$ and subtracted as background. 
However, the fraction, $N_{B\to J/\psi\to\mu}/N_{\rm HF}$, is quite 
small based on the measurements of the $B\to\jpsi$ 
fraction~\cite{Aidala:2017yte}.

\begin{figure}[thb]
\includegraphics[width=1.0\linewidth]{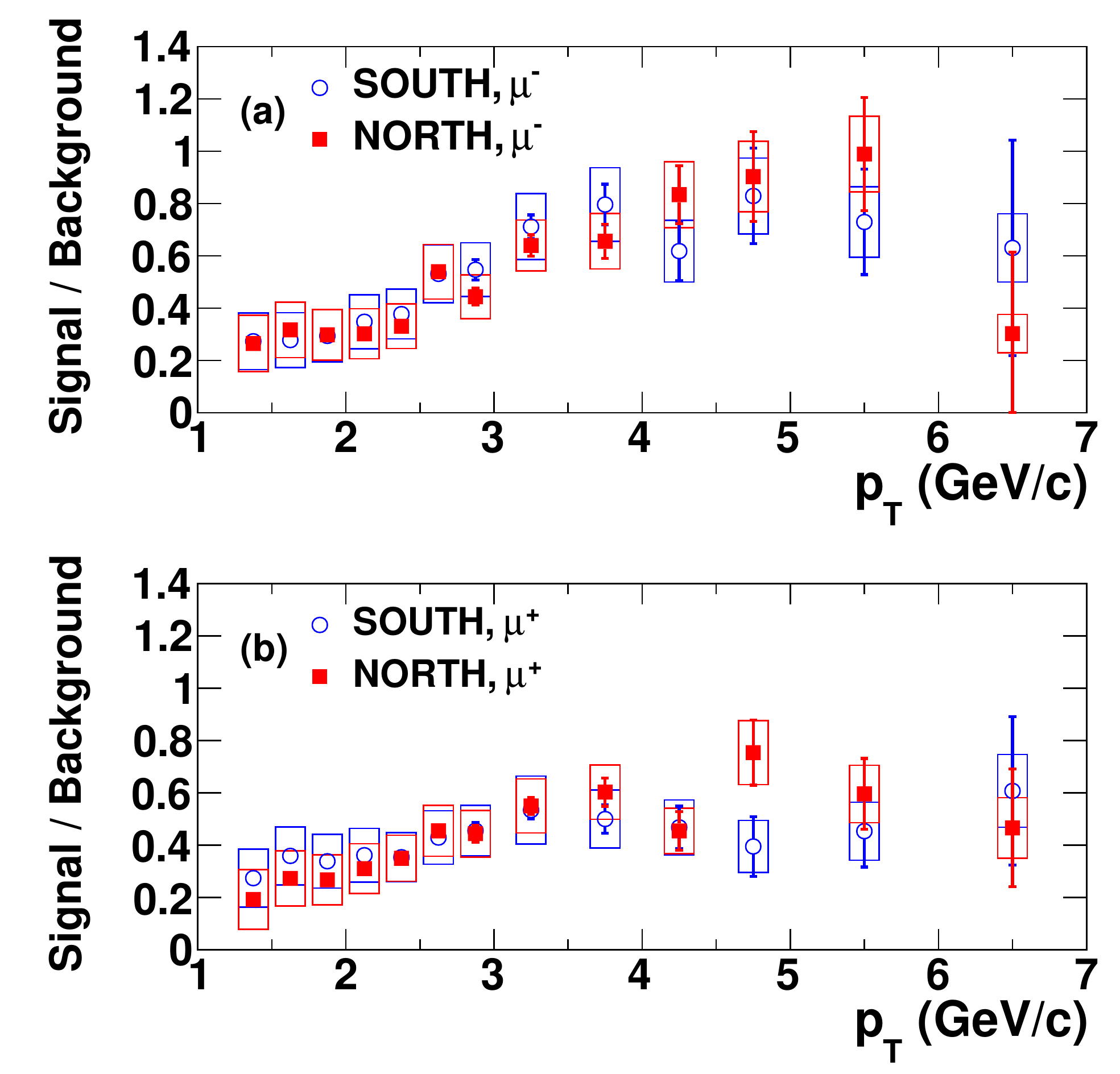}
\caption{\label{fig:SBratio}
Signal-to-background ratio of (a) negatively-charged and 
(b) positively-charged tracks.  Each panel includes results in the 
North (closed [red] rectangle) and South (open [blue] circle) arms. 
Vertical bars (boxes) correspond to the statistical (systematic) 
uncertainties.}
\end{figure}

Figure~\ref{fig:gap4_comp} shows the \pt spectra of inclusive muon 
tracks and estimated background components; the relative contribution 
from each source varies with \pt. After subtraction of backgrounds from 
light hadrons and \jpsi, the \pt spectra of muons from open heavy-flavor 
decays can be obtained. Figure~\ref{fig:SBratio} shows the 
signal-to-background ratio 
($\frac{N_{\rm HF}}{N_{\rm DM} + N_{\rm PH} + N_{J/\psi\to\mu}}$) 
of negatively (top panel) and positively (bottom 
panel) charged tracks; blue open circle (red closed rectangle) points 
represent the results in the South (North) arm. Vertical bars (boxes) 
around the data points are statistical (systematic) uncertainties; 
details on systematic uncertainties will be described in the following 
section.   Because $K^{+}$ has a longer nuclear interaction length than 
other light hadrons, the signal-to-background ratio of 
positively-charged tracks is smaller than that of negatively-charged 
tracks.

\subsubsection{Acceptance and efficiency correction}
\label{sec:acceff}

The acceptance and efficiency correction is evaluated by using a 
single-muon simulation. The same simulation procedure as for the 
hadron-cocktail simulation is used, and reconstructed muons are filtered 
with the same track quality cuts and fiducial cuts as was applied to the 
data.  Because detector performance throughout the data-taking period is 
stable, one reference run is used to calculate the correction factors. 
The variation of the number of muon candidates per event throughout the 
data-taking period is 8.1\% (4.6\%) for the South (North) arm, and the 
quadratic sum with the systematic uncertainty on the MuTr (4\%) and MuID 
(2\%) is assigned to the systematic uncertainty on the acceptance and 
efficiency correction.

\subsubsection{Systematic uncertainty}

There are three major sources of systematic uncertainty; the background 
estimation ($\delta_{bkg}$), the acceptance and efficiency correction 
($\delta_{A\varepsilon}$), and the BBC efficiency ($\delta_{\rm BBC}$).

The sources of $\delta_{bkg}$ are listed here:
\begin{description}

\item[$\delta_{\rm trig}$] A 5\% (15\%) systematic uncertainty is 
assigned to the MuID trigger efficiency for tracks at MuID Gap4 (Gap3) 
by considering the statistical uncertainty of tracks in the non-MuID 
triggered events, and the uncertainty is included in the systematic 
uncertainty on the $N_{\rm DM}$ (Gap4) and $N_{\rm PH}$ (Gap3).

\item[$\delta_{\rm sim}$] The hadron-cocktail simulation with the thick 
absorber ($\sim 13\lambda_{I}$) can be a source of systematic 
uncertainty. In case of the $N_{\rm DM}$ in $\pt<3~{\rm GeV}/c$ where 
background can be constrained with muons, a 10\% systematic uncertainty 
is assigned conservatively due to extraction of the slope in the 
normalized $z_{\rm vtx}$ distributions. The difference between the two 
methods of tuning described in Sec.~\ref{sec:bkg_estimation} is assigned 
to the systematic uncertainty on the $N_{\rm DM}$ in $\pt>3~{\rm GeV}/c$ 
and the $N_{\rm PH}$. The systematic uncertainty on the $N_{\rm DM}$ 
($N_{\rm PH}$) is 10--15\% (10--40\%) depending on \pt.

\item[$\delta_{\rm input}$]  Because there is no precise measurement of 
$\pi^{\pm}$ and $K^{\pm}$ production at forward rapidity, a 30\% 
systematic uncertainty is assigned to the estimation of $K/\pi$ ratio 
based on the systematic uncertainty of measurements at 
midrapidity~\cite{Adare:2011vy,Agakishiev:2011dc}. The impact on 
$N_{\rm HF}$ is evaluated by performing the hadron-cocktail tuning 
procedure with various initial $K/\pi$ ratios, and the variation of 
$N_{\rm HF}$ is less than 10\%. The uncertainty on the shape of the \pt 
distribution is negligible, because the tuning of the hadron-cocktail 
simulation can take into account a \pt dependence. A 10\% systematic 
uncertainty is assigned to $N_{\rm HF}$ conservatively.

\item[$\delta_{J/\psi\to\mu}$] The upper and lower limit of systematic 
uncertainty on the \jpsi cross section measurement is taken into account 
for the systematic uncertainty on $N_{J/\psi\to\mu}$. The contribution 
from $B$ decays is also considered. A 3\% systematic uncertainty is 
assigned to the $N_{\rm HF}$ due to the uncertainty on the 
$N_{J/\psi\to\mu}$.

\end{description}

For the systematic uncertainty on the $N_{\rm HF}$, the $\delta_{\rm trig}$ 
and $\delta_{\rm sim}$ on the $N_{\rm DM}$ ($N_{\rm PH}$) are 
propagated into the $N_{\rm HF}$ with the ratio of $N_{\rm DM}/N_{\rm HF}$ 
($N_{\rm PH}/N_{\rm HF}$). This propagated uncertainty is combined 
with the $\delta_{\rm input}$ and $\delta_{J/\psi\to\mu}$ on the $N_{\rm HF}$ 
as a quadratic sum. The $\delta_{bkg}$ is 8--40\%, depending on \pt.

There are also systematic uncertainties on the acceptance and efficiency 
correction ($\delta_{A\varepsilon}$) and the BBC efficiency 
($\delta_{\rm BBC}$); see the discussion in~\cite{Adler:2003pb}. For the 
$\delta_{A\varepsilon}$, all sources described in Sec.~\ref{sec:acceff} 
are added in quadrature, and 9.3\% and 6.4\% systematic uncertainties 
are assigned to the South and North arm, respectively.

Table~\ref{tab:sys_xection} summarizes the systematic uncertainty on the 
cross section of muons from open heavy-flavor decays, and the quadratic 
sum of the three components is the final systematic uncertainty.

\begin{table}[tbh]
\caption{\label{tab:sys_xection}
Summary of systematic uncertainties on the cross section of muons from 
open heavy-flavor decays.}
\begin{ruledtabular}
\begin{tabular}{ccc}
& Component & Value \\\hline
$\delta_{bkg}$ & background estimation & 8--40\%, varies with \pt\\
$\delta_{A\varepsilon}$ & Acceptance and efficiency & 9.3\%(S), 6.4\%(N)\\
$\delta_{\rm BBC}$ & BBC efficiency & 10.1\%\\ 
\\
 & sum & 17--43\%, varies with \pt
\end{tabular}
\end{ruledtabular}
\end{table}

\subsection{Transverse Single-Spin Asymmetry}

\subsubsection{Determination of the TSSA}

Both of the proton beams are transversely polarized 
at the interaction point.  The TSSA ($A_N$) in the yield of muons from 
heavy-flavor decays is obtained for each beam separately by summing 
over the spin information of the other beam.  The final asymmetry is 
calculated as the weighted average of the asymmetries for the two beams.

The maximum likelihood method is used for this measurement. The 
likelihood $\mathcal{L}$ is defined as,

\begin{equation}
\mathcal{L} = \prod (1 + P \cdot A_N \sin(\phi_{\rm pol} - \phi_i)),
\end{equation}
where $P$ is the polarization, $\phi_{\rm pol}$ is the direction of beam 
polarization ($+\frac{\pi}{2}$ or $-\frac{\pi}{2}$), and $\phi_i$ is the 
azimuthal angle of each track in the PHENIX lab frame. The unbinned 
likelihood method is used in this study, so that the result is not 
biased by low statistics bins. The likelihood function is usually 
written in logarithmic form

\begin{equation}
\label{eq:maxlikelihood}
\log \mathcal{L} = \sum \log(1 + P \cdot A_N \sin(\phi_{\rm pol} - \phi_i)),
\end{equation}
The $A_N$ value is determined by maximizing $\log \mathcal{L}$. The 
statistical uncertainty of the log-likelihood estimator is related to 
its second derivative,

\begin{equation}
\label{eq:maxlikelihood_err}
\sigma^2(A_N) = (-\frac{\partial^2 \mathcal{L}}{\partial A_N^2})^{-1}.
\end{equation}

\subsubsection{Inclusive- and background-asymmetry estimation}

We study tracks that penetrate to the last MuID gap (Gap4); these tracks 
are created by muons from open heavy-flavor decays, punch-through 
hadrons, muons from light hadrons, and muons from \jpsi decay. The 
contribution from other sources is negligible as discussed in 
Sec.~\ref{sec:bkg_estimation}. To obtain the asymmetry of muons 
from open heavy-flavor decays ($A_N^{\rm HF}$), the asymmetry of the 
background from light hadrons ($A_N^{\rm h}$) and muons from \jpsi 
($A_N^{J/\psi\to\mu}$) should be eliminated from the asymmetry of 
inclusive muon candidates ($A_N^{\rm incl}$).  Because hadron tracks can be 
selected with the $p_{z}$ cut, $A_N^{\rm h}$ is obtained from the 
asymmetry of stopped hadrons at MuID Gap3. Possible differences between 
the $A_N$ of stopped hadrons at MuID Gap3 and the mixture of decay muons 
and punch-through hadrons at MuID Gap4 is studied with the 
hadron-cocktail simulation. The details are described in 
Sec.~\ref{sec:an_sys}.

For the estimation of $A_N^{J/\psi\rightarrow\mu}$, a previous PHENIX 
$A_N^{J/\psi}$ measurement~\cite{Adare:2010bd} is used. The asymmetry of 
single muons from \jpsi decay ($A_N^{J/\psi\rightarrow\mu}$) is 
estimated from a decay simulation with the initial $A_N^{J/\psi}$ 
in~\cite{Adare:2010bd} ($A_N^{J/\psi}=-0.002\pm0.026$ at $x_F<0$, and 
$-0.026\pm0.026$ at $x_F>0$). The initial \pt and rapidity distributions 
of \jpsi are taken from~\cite{Adare:2011vq}. The obtained 
$A_N^{J/\psi\rightarrow\mu}$ is $-0.002^{+0.018}_{-0.022}$ at $x_F<0$ 
and $-0.019^{+0.019}_{-0.025}$ at $x_F>0$. A possible effect from 
$J/\psi$ polarization is tested by assuming maximum polarization, and 
the variation of $A_N^{J/\psi\rightarrow\mu}$ is $<0.001$.  Because the 
variation due to $J/\psi$ polarization is much smaller than the 
variation from the uncertainty of $A_N^{J/\psi}$, the $J/\psi$ 
polarization effect is not included to evaluate 
$A_N^{J/\psi\rightarrow\mu}$ and the systematic uncertainty.

Once $A_N^{\rm h}$ and $A_N^{J/\psi\to\mu}$ are determined, the $A_N$ of 
muons from open heavy-flavor decays and its uncertainty can be obtained as

\begin{equation}
\label{eq:aneq}
A_N^{\rm HF}=\frac{A_N^{\rm incl}-f_{\rm h}\cdot A_N^{\rm h}-f_{J/\psi}\cdot A_N^{J/\psi\to\mu}}{1-f_{\rm h}-f_{J/\psi}},
\end{equation}

\begin{equation}
\label{eq:anerreq}
\delta A_N^{\rm HF}=\frac{\sqrt{(\delta
A_N^{\rm incl})^2+f_{\rm h}^2\cdot (\delta A_N^{\rm h})^2+f_{J/\psi}^2\cdot (\delta A_N^{J/\psi\to\mu})^2}}
{1-f_{\rm h}-f_{J/\psi}},
\end{equation}
where $f_{\rm h}=(N_{\rm DM}+N_{\rm PH})/N_{\rm incl}$ is the fraction 
of the light-hadron background, and $f_{J/\psi}=N_{J/\psi\to\mu}/N_{\rm 
incl}$ is the fraction of muons from \jpsi. Both fractions ($f_{\rm h}$ 
and $f_{J/\psi}$) are determined from the background estimation 
described above. $\delta A_N^{J/\psi\to\mu}$, estimated from the 
previous PHENIX measurement, is included in the systematic uncertainty.

\subsubsection{Systematic Uncertainty}
\label{sec:an_sys}

The systematic uncertainty is determined from variation of $A_N^{\rm HF}$ 
between the upper and lower limit of each background source. An 
additional systematic uncertainty is derived from the comparison between 
the two $A_N^{\rm HF}$ calculation methods; the maximum likelihood 
method (Eq.~\eqref{eq:maxlikelihood}) and the polarization formula 
(Eq.~\eqref{eq:anvsphi_pp}). The final systematic uncertainty is 
calculated as the quadratic sum of systematic uncertainties from each 
source ($\delta{A_N^{\delta f_{\rm h}}}$, $\delta{A_N^{\rm h}}$, 
$\delta{A_N^{J/\psi\to\mu}}$, and $\delta{A_N^{\rm method}}$), described 
here:

\begin{description}

\item[$\delta{A_N^{\delta f_{\rm h}}}$] Systematic uncertainty on the 
fraction of light-hadron background ($\delta f_{\rm h}$) from 
Fig.~\ref{fig:SBratio} is an important source of systematic uncertainty 
on $A_{N}^{\rm HF}$. The upper and lower limits of $A_N^{\rm HF}$ are 
calculated using Eq.~\eqref{eq:aneq} with the upper and lower limits of 
the fraction of the light-hadron background ($f_{\rm h}\pm\delta f_{\rm h}$).

\item[$\delta{A_N^{\rm h}}$] The asymmetry of the light-hadron 
background ($A_N^{\rm h}$) at MuID Gap4 is estimated by using stopped 
hadrons at MuID Gap3. Due to decay kinematics, the $A_N^{\rm h}$ at MuID 
Gap4 can be different from the $A_N^{\rm h}$ measured at MuID Gap3. In 
order to quantify the difference, a simulation study using the decay 
kinematics of light hadrons from the hadron-cocktail in 
Sec.~\ref{sec:bkg_estimation} and an input asymmetry ($A_N^{\rm input}$) 
is performed. $A_N^{\rm input}$ is taken as $0.02\times\pt$ (with \pt in 
GeV$/c$) at $\pt<5~{\rm GeV}/c$ and 0.1 at $\pt>5~{\rm GeV}/c$, based on 
the most extreme case of $A_N^{\rm h}$ measured at MuID Gap3. The 
detailed procedure is as follows:

\begin{enumerate}

\item Generate a random spin direction ($\uparrow$,$\downarrow$) for all 
tracks.

\item Apply a weight ($1\pm A_N^{\rm input}\cdot \cos\phi_0$) for each 
track based on the manually assigned initial asymmetry ($A_{N}^{\rm 
input}$). The sign is determined from the random polarization direction 
in step 1, and $\phi_{0}$ is the azimuthal angle of the track at the 
generation level.

\item Extract $A_N^{\rm reco}$ of the tracks at MuID Gap3 and Gap4 with 
the azimuthal angle and momentum of the reconstructed tracks by fitting 
the asymmetry of the two polarization cases with $A_N^{\rm reco}\cdot 
\cos\phi_{0}$.

\end{enumerate}

The largest difference between $A_N^{\rm reco}$ at MuID Gap3 and Gap4 is 
$\sim 0.008$ in the entire \pt range, so $\pm0.008$ is assigned to the 
systematic uncertainty. In the case of $x_F$ binning, the difference of 
$A_N^{\rm reco}$ at MuID Gap3 and Gap4 is quite small ($<0.001$).

\item[$\delta{A_N^{J/\psi\to\mu}}$] 
The systematic uncertainty from $A_N^{J/\psi\to\mu}$ is determined from 
the $J/\psi\to\mu$ simulation with the upper and lower limits of 
$A_N^{J/\psi}$ in~\cite{Adare:2010bd}. Propagation to $A_N^{\rm HF}$ is 
calculated using Eq.~\eqref{eq:aneq}. The effect from $B\to J/\psi$ is 
negligible due to its small fraction in the inclusive \jpsi.

\item[$\delta{A_N^{\rm method}}$]
The $A_N^{\rm incl}$ results from the maximum likelihood method at 
Eq.~\eqref{eq:maxlikelihood} are compared with result using the 
polarization formula at Eq.~\eqref{eq:anvsphi_pp}.  Because the measurement 
of $A_N^{\rm h}$ using tracks at MuID Gap3 suffer from large statistical 
fluctuations, the difference of two methods with inclusive tracks at 
MuID Gap4 is used for both $A_N^{\rm incl}$ and $A_N^{\rm h}$ variations 
using Eq.~\eqref{eq:aneq}. $A_N(\phi)$ of inclusive tracks for each \pt 
or $x_F$ bin is calculated as,

\begin{equation}
\label{eq:anvsphi_pp}
\ A_N(\phi)=\frac{\sigma^{\uparrow}(\phi)-\sigma^{\downarrow }(\phi)}{\sigma^{\uparrow}(\phi)+\sigma^{\downarrow}(\phi)}\\
=\frac{1}{P}\cdot\frac{N^{\uparrow}(\phi)-R\cdot N^{\downarrow}(\phi)}{N^{\uparrow}(\phi)+R\cdot N^{\downarrow}(\phi)},                              \\
\end{equation}
where $P$ is the average beam polarization, $\sigma^{\uparrow }$, 
$\sigma^{\downarrow}$ are cross sections for each polarization, 
$N^{\uparrow}$, $N^{\downarrow}$ are yields for two polarizations and 
$R = L^{\uparrow}/L^{\downarrow}$ is the relative luminosity where the 
luminosity ($L^{\uparrow}, L^{\downarrow}$) is measured by the BBC 
detectors. $A_N^{\rm incl}$ is calculated by fitting the $A_N(\phi)$ 
distribution with a function $\pm A_{N} \cdot \cos\phi$, where $\pm$ 
depends on the beam direction. The systematic uncertainty on $A_N^{\rm HF}$ 
is evaluated by propagating variations of $A_N^{\rm incl}$ and 
$A_N^{\rm h}$ between the maximum likelihood method and the polarization 
formula.

\end{description}

\section{Results}
\label{sec:results}

\subsection{Cross section of muons from open heavy-flavor decays}

The invariant cross section of muons from open heavy-flavor decays is 
calculated as

\begin{equation}
E\frac{d^{3}\sigma}{dp^{3}}=\frac{1}{2\pi\pt\Delta\pt\Delta y}
\frac{(N_{\rm HF}/\varepsilon_{\rm BBC}^{{\rm HF}})\cdot\sigma_{pp}^{\rm inel}}{(N_{\rm evt}/\varepsilon_{\rm BBC}^{\rm MB})\cdot A\varepsilon},
\end{equation}
where $\Delta\pt$ and $\Delta y$ are the bin widths in \pt and $y$, 
$N_{\rm evt}$ is the number of sampled MB events, $\varepsilon_{\rm 
BBC}^{\rm MB}$ ($\varepsilon_{\rm BBC}^{\rm HF}$) is the BBC correction 
factor for the trigger efficiency of MB events (events containing muons 
from open heavy-flavor decays), $A\varepsilon$ is the detector 
acceptance and track reconstruction efficiency, and $\sigma_{pp}^{\rm 
inel}=42\pm3~{\rm mb}$ is the inelastic cross section of \pp collisions 
at $\sqrts=200~{\rm GeV}$.

\begin{figure}[thb]
\includegraphics[width=1.0\linewidth]{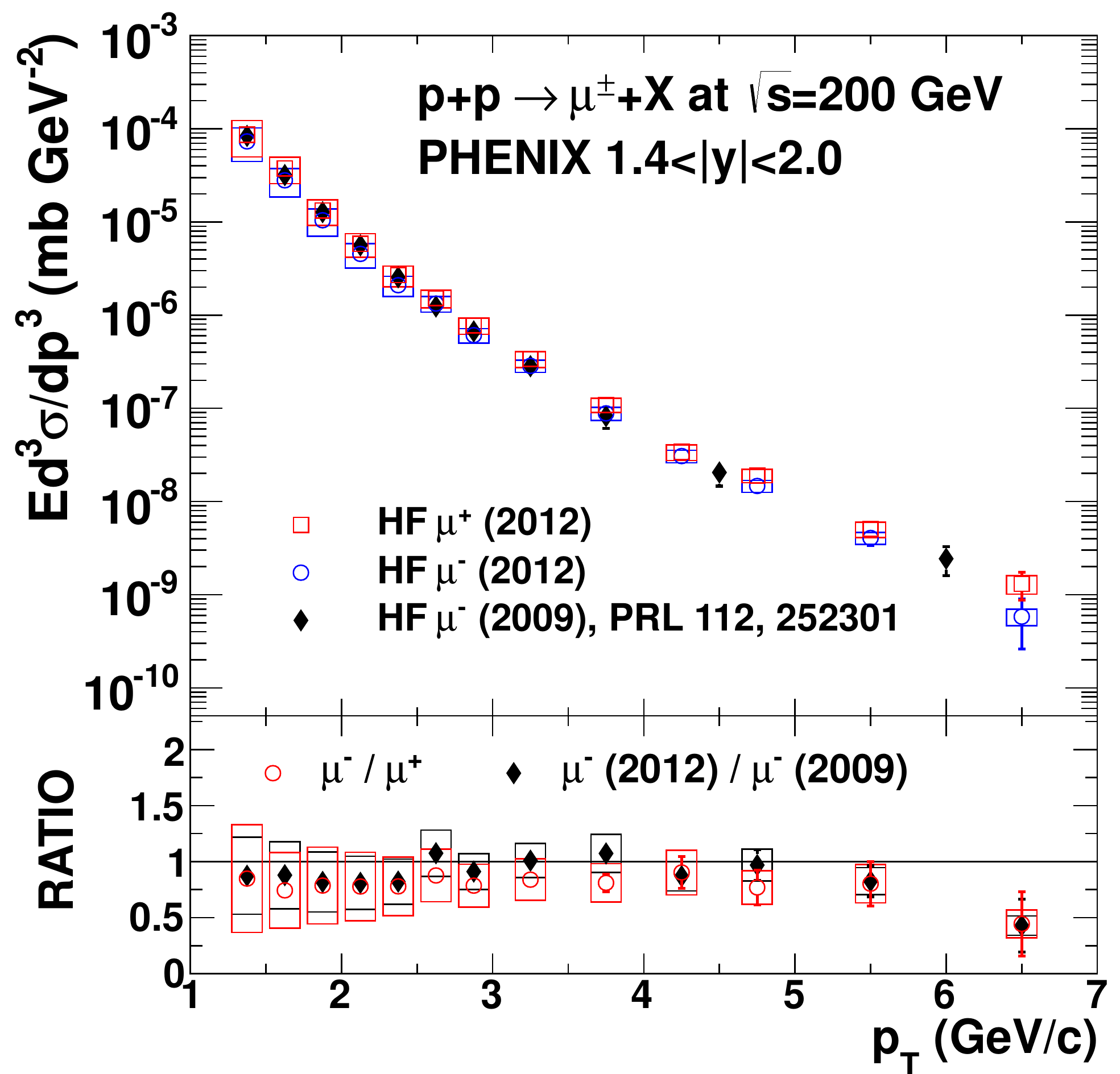}
\caption{\label{fig:HFmu_charge} 
(top) Invariant cross section of muons from open heavy-flavor decays as 
a function of \pt in \pp collisions at $\sqrts=200~{\rm GeV}$ at forward 
rapidity. (bottom) Ratio of invariant cross sections. Vertical bars 
(boxes) correspond to the statistical (systematic) uncertainties.}
\end{figure}

Figure~\ref{fig:HFmu_charge} shows the invariant cross section of 
positively- (open square) and negatively-charged (open circle), muons 
from open heavy-flavor decays as a function of \pt in \pp collisions at 
$\sqrts=200~{\rm GeV}$. Vertical bars (boxes) correspond to the 
statistical (systematic) uncertainties. The previous PHENIX results for 
negatively charged muons~\cite{Adare:2013lkk} are shown and vertical 
bars represent total uncertainties. The bottom panel shows the ratio 
between positively- and negatively-charged muons from open heavy-flavor 
decays (red open circles); the two \pt spectra are consistent within the 
systematic uncertainties which are dominated by the uncertainty from the 
hadron contamination. The comparison with the previous PHENIX results 
for negative muons is also presented as a ratio (black diamonds); the 
fit function in~\cite{Adare:2013lkk} is used to make a ratio at 
$\pt>4.0~{\rm GeV}/c$. The uncertainties from the new results are 
included in the ratio, and two results are in good agreement.

\subsection{Transverse single-spin asymmetry}

\begin{figure}[thb]
\centering
\includegraphics[width=1.0\linewidth]{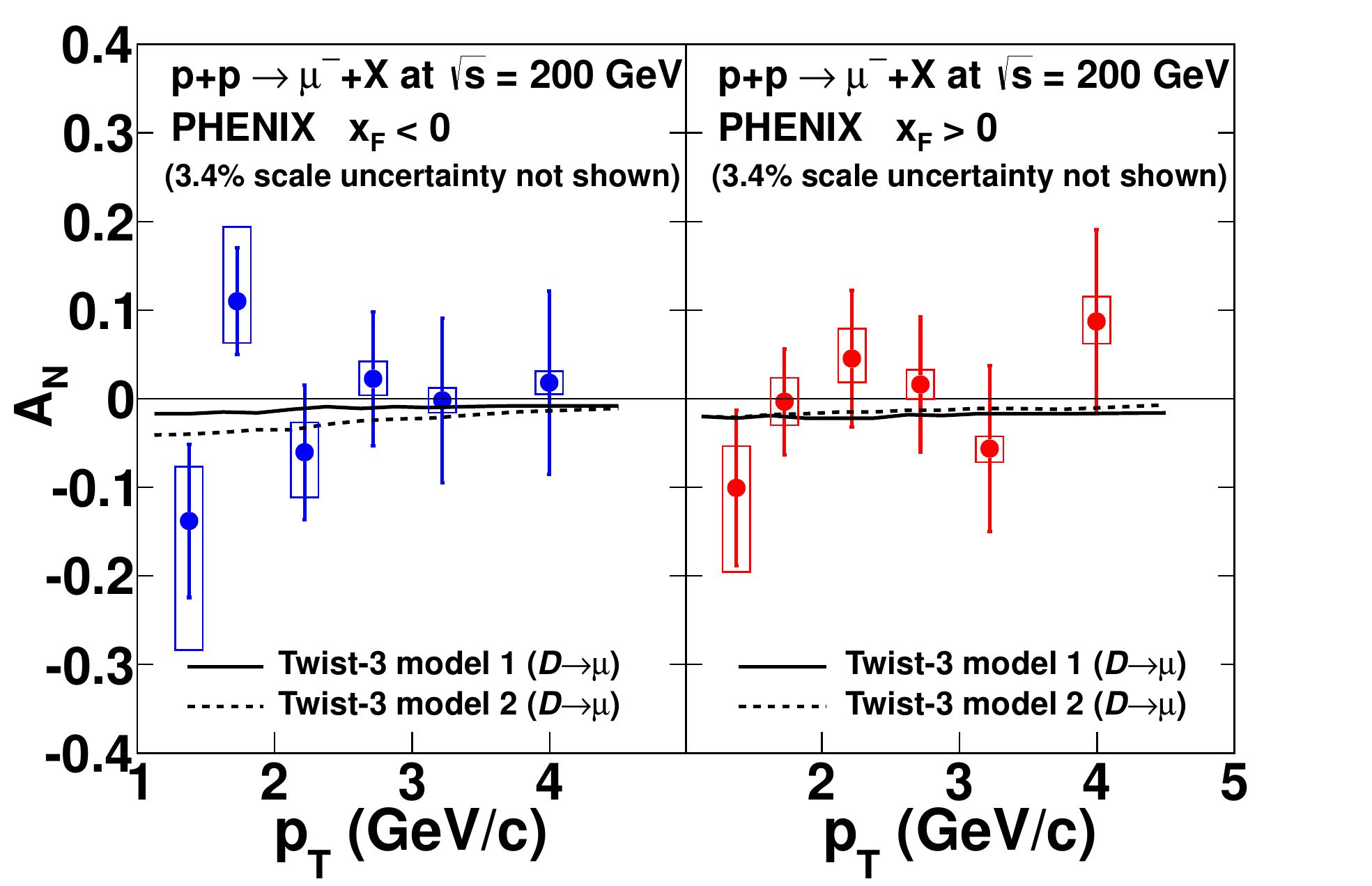}
\caption{\label{fig:pt_AN_minus}
$A_N$ of negatively-charged muons from open heavy-flavor decays as a 
function of $p_T$ in the backward ($x_F<0$, left) and forward ($x_F>0$, 
right) regions. Vertical bars (boxes) represent statistical (systematic) 
uncertainties. Solid and dashed lines represent twist-3 model 
calculations~\protect\cite{Koike:2011mb}, described in 
Sec.~\ref{sec:discussion}.}
\end{figure}

\begin{figure}[thb]
\centering
\includegraphics[width=1.0\linewidth]{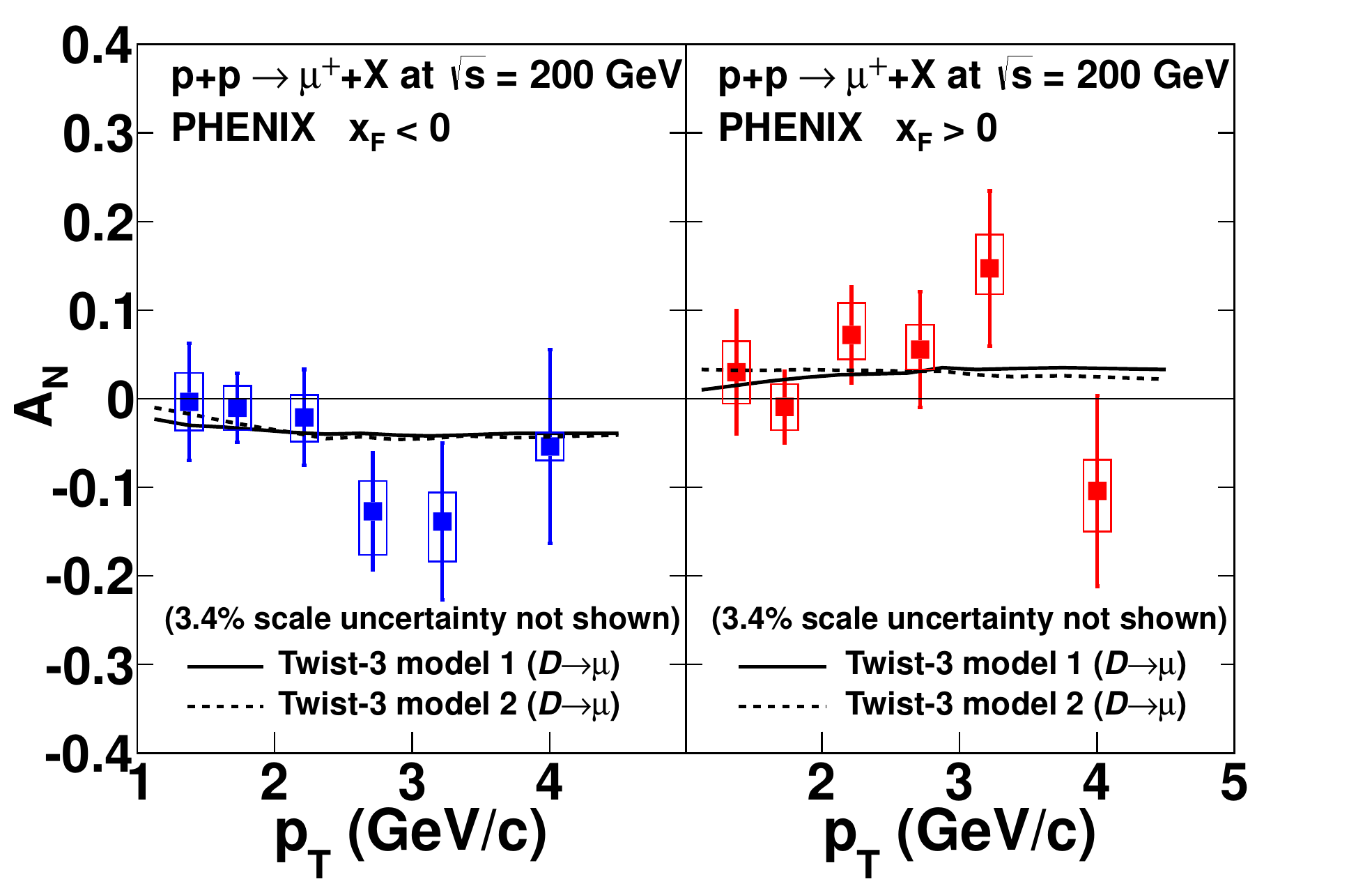}
\caption{\label{fig:pt_AN_plus}
$A_N$ of positively-charged muons from open heavy-flavor decays as a 
function of $p_T$ in the backward ($x_F<0$, left) and forward ($x_F>0$, 
right) regions. Vertical bars (boxes) represent statistical (systematic) 
uncertainties. Solid and dashed lines represent twist-3 model 
calculations~\protect\cite{Koike:2011mb}, described in 
Sec.~\ref{sec:discussion}.}
\end{figure}

\begin{figure}[thb]
\centering
\includegraphics[width=1.0\linewidth]{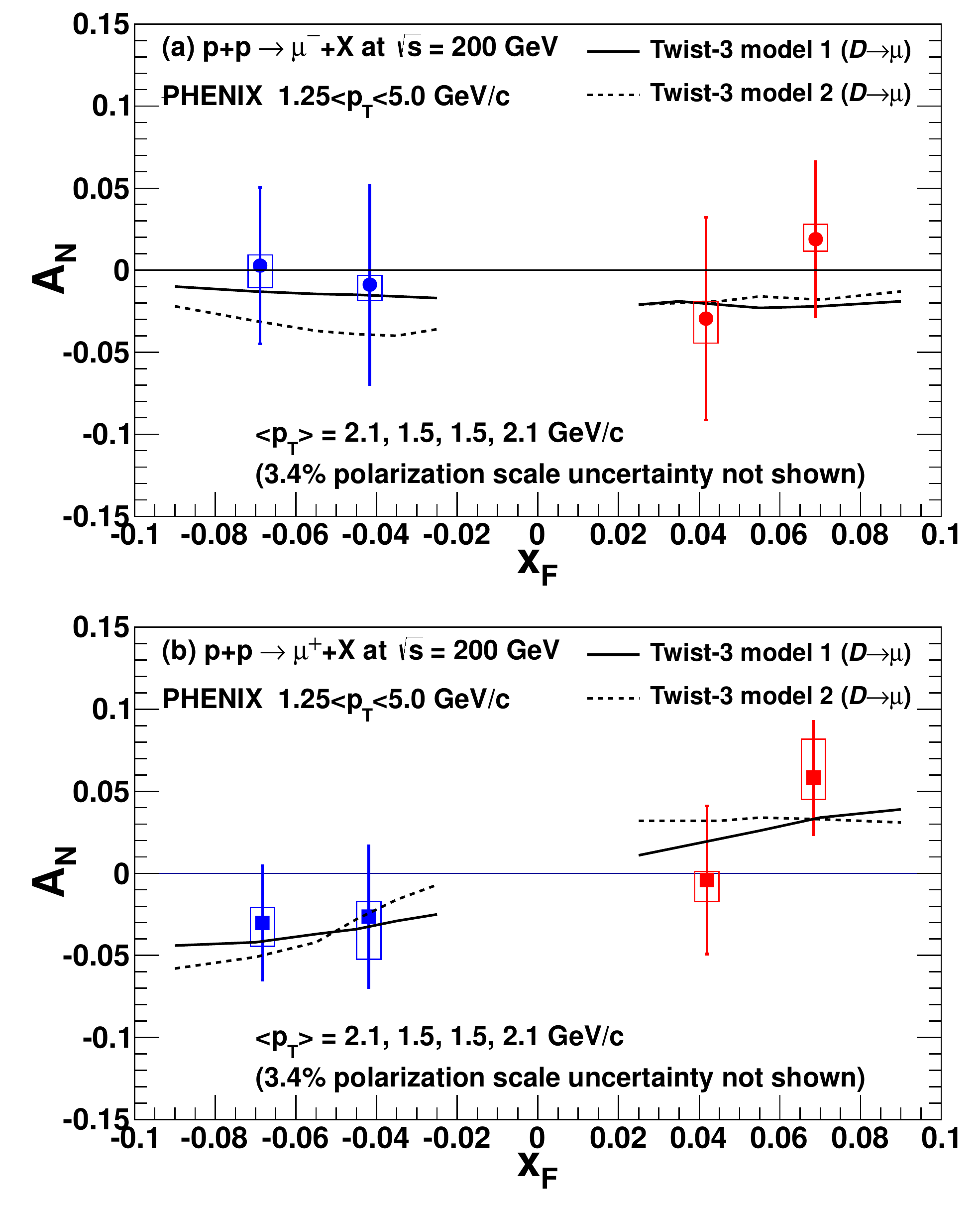}
\caption{\label{fig:xf_AN_bothchg}
$A_N$ of (a) negatively-charged and (b) positively-charged muons 
from open heavy-flavor decays as a function of $x_F$, where $x_F > 0$ is 
along the direction of the polarized proton. Vertical bars (boxes) 
represent statistical (systematic) uncertainties. Solid and dashed lines 
represent twist-3 model calculations~\protect\cite{Koike:2011mb}, described in 
Sec.~\protect\ref{sec:discussion}.}
\end{figure}

The TSSA of muons from open heavy-flavor decays is calculated by using 
Eq.~\eqref{eq:aneq} and the statistical uncertainty is determined by 
using Eq.~\eqref{eq:anerreq}. Figures~\ref{fig:pt_AN_minus} 
and~\ref{fig:pt_AN_plus} present the TSSA of negatively- ($A_N^{\mu^-}$) 
and positively- ($A_N^{\mu^+}$) charged muons from open heavy-flavor as 
a function of $p_T$ in the forward ($x_F>0$) and backward ($x_F<0$) 
regions with respect to the polarized-proton beam direction. 
Figure~\ref{fig:xf_AN_bothchg} shows the TSSA versus $x_F$ of muons from 
open heavy-flavor decays. Vertical bars (boxes) represent statistical 
(systematic) uncertainties; a scale uncertainty from the polarization 
(3.4\%) is not included. $A_N^{\mu^+}$ in the negative $x_F$ region, 
shown in the left panel of Fig.~\ref{fig:pt_AN_plus}, shows some 
indication of a negative asymmetry; in the combined \pt range of 
$2.5<\pt<5.0~{\rm GeV}/c$ the asymmetry is $-0.117\pm0.048{\rm 
(stat)}\pm0.037{\rm (syst)}$. However, the combined asymmetries for all 
\pt or $x_F$ bins are consistent with zero within total uncertainties. 
Other results for $A_N^{\mu^+}$ at positive $x_F$ and $A_N^{\mu^-}$ in 
all kinematic regions are consistent with zero within statistical 
uncertainties. The results are tabulated in Tables~\ref{tab:AN_PT} 
and~\ref{tab:AN_XF}, while Tables~\ref{tab:AN_PT_SYS} 
and~\ref{tab:AN_XF_SYS}, list the systematic uncertainties from each 
source.

\section{Discussion}
\label{sec:discussion}

\begin{figure}[thb]
\includegraphics[width=1.0\linewidth]{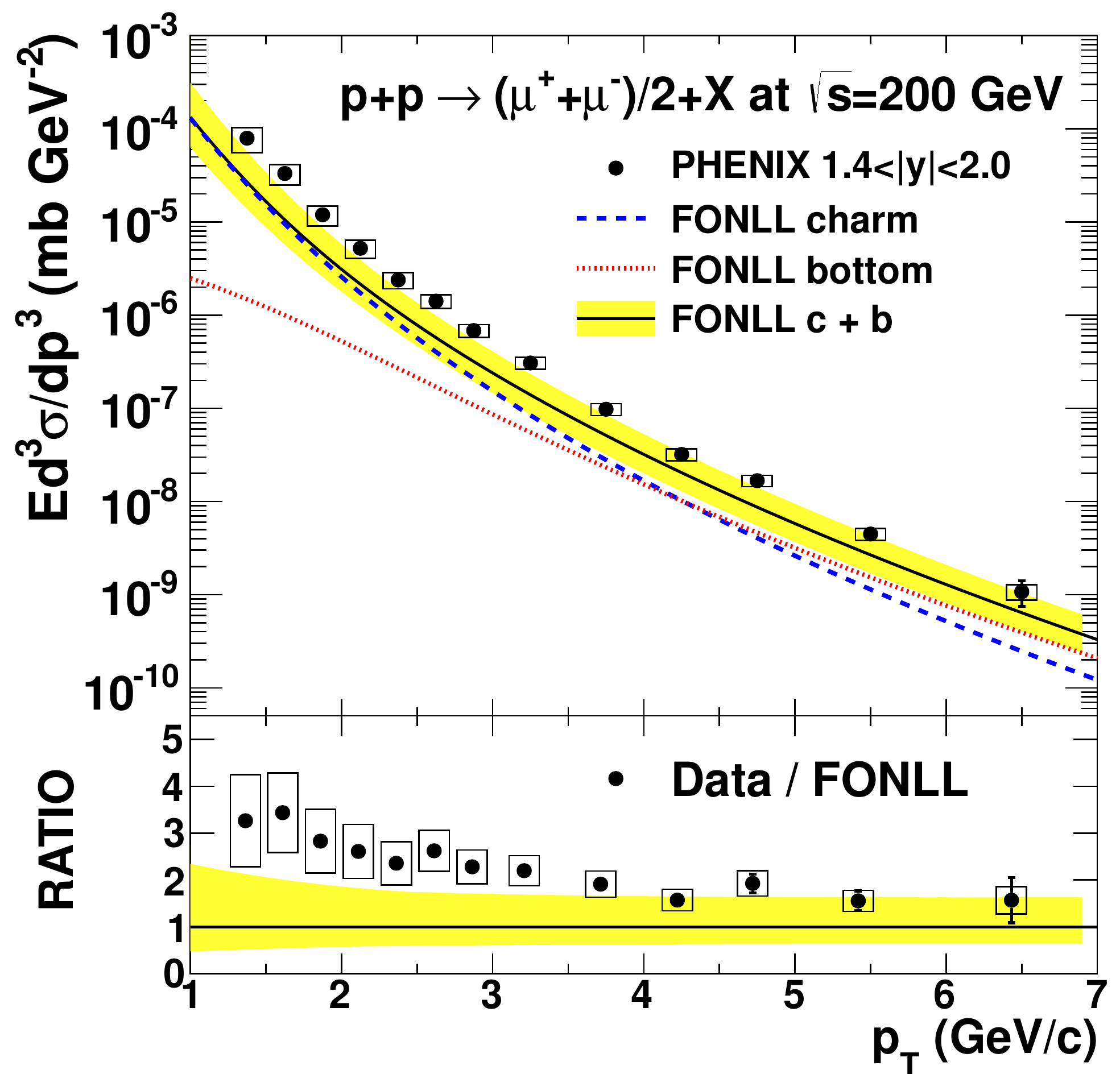}
\caption{\label{fig:HFmu_combined} 
(top) Charge-combined, invariant cross section of muons from open 
heavy-flavor decays as a function of \pt in \pp collisions at 
$\sqrts=200~{\rm GeV}$ at forward rapidity. The solid line and band 
represent the FONLL calculation for charm and bottom and its systematic 
uncertainty. The dashed and dotted curves show contributions from charm 
and bottom separately. (bottom) Ratio between the data and the FONLL 
calculation. Vertical lines (boxes) represent statistical (systematic) 
uncertainties of the data.}
\end{figure}

Figure~\ref{fig:HFmu_combined} shows the charge-combined, invariant 
cross section of muons from open heavy-flavor decays as a function of 
\pt. Vertical bars (boxes) correspond to the statistical (systematic) 
uncertainties. The solid line in Fig.~\ref{fig:HFmu_combined} represents 
the fixed-order-plus-next-to-leading-log (FONLL) calculation of muons 
from open heavy-flavor decays from charm and 
bottom~\cite{Cacciari:1998it}, and the band around the line represents 
the systematic uncertainty from the renormalization scale, factorization 
scale, and heavy ($c$ and $b$) quark masses. The bottom panel shows the 
ratio between the data and the FONLL calculation. In general, the 
agreement between the data and the FONLL prediction becomes better with 
increasing \pt where the systematic uncertainties of both are 
decreasing. At $\pt<4~{\rm GeV}/c$ where the charm contribution is 
larger than that from bottom, the measured yield is larger than the 
FONLL calculation, but systematic uncertainties are large in both the 
data and the theoretical calculation. Recently, a theoretical approach 
within the gluon saturation (Color-Glass-Condensate) framework also 
presents the cross section of leptons from heavy-flavor decays in \pp 
and $p$+$A$ collisions~\cite{Fujii:2015lld}.

A recent theoretical calculation~\cite{Koike:2011mb} incorporating the 
collinear factorization framework makes predictions for $A_N$ in the 
production of $D$-mesons ($A_N^{D}$) produced by the gluon-fusion 
($gg\rightarrow c\bar{c}$) process and therefore is sensitive to the 
trigluon correlation functions which depend on the momentum fraction of 
the gluon in the proton in the infinite-momentum frame ($x$-Bjorken). 
Two model calculations, assuming either a linear $x$-dependence (Model 1 
in Fig.~\ref{fig:pt_AN_minus},~\ref{fig:pt_AN_plus}, 
and~\ref{fig:xf_AN_bothchg}) or a $\sqrt{x}$-dependence (Model 2 in 
Fig.~\ref{fig:pt_AN_minus},~\ref{fig:pt_AN_plus}, 
and~\ref{fig:xf_AN_bothchg}), for the nonperturbative functions 
participating in the twist-3 cross section for $A_N^{D}$ are introduced 
to compare their behavior in the small-$x$ region, and the overall 
$A_N^{D}$ scale is determined by assuming $|A_N^D|\leq0.05$ at 
$|x_F|<0.1$.

To compare with our results for $A_N^{\mu}$, the decay 
kinematics and cross section of $D\to\mu$ from 
{\sc pythia}~\cite{Sjostrand:2014zea} have been used to convert 
$A_N^{D}$ into $A_N^{\mu}$. The theory calculations of the $x_F$ and \pt 
dependence of $A_N$ for $D^0$, $\bar{D^0}$, $D^+$, and $D^-$ at 
$-0.6<x_F^{D}<0.6$ and $1<p_{T}^{D}<10~{\rm GeV}/c$ are used as the 
input $A_N^{D}$ to the simulation. A similar procedure to that described 
in the systematic-uncertainty evaluation for $\delta A_N^{\rm h}$ is 
used. A weight of ($1\pm A_N^{D}(p_T^D,x_F^D)\cdot \sin(\phi^D-\phi_{\rm 
pol})$) is applied for each muon from a $D$ meson and the sign is 
determined with a random polarization direction 
($\uparrow$,$\downarrow$). Then, $A_N^{\mu}$ is extracted by fitting the 
asymmetry of the two polarization cases with $A_N^{\mu}\cdot\cos\phi^{\mu}$.

\begin{figure}[thb]
\centering
\includegraphics[width=1.0\linewidth]{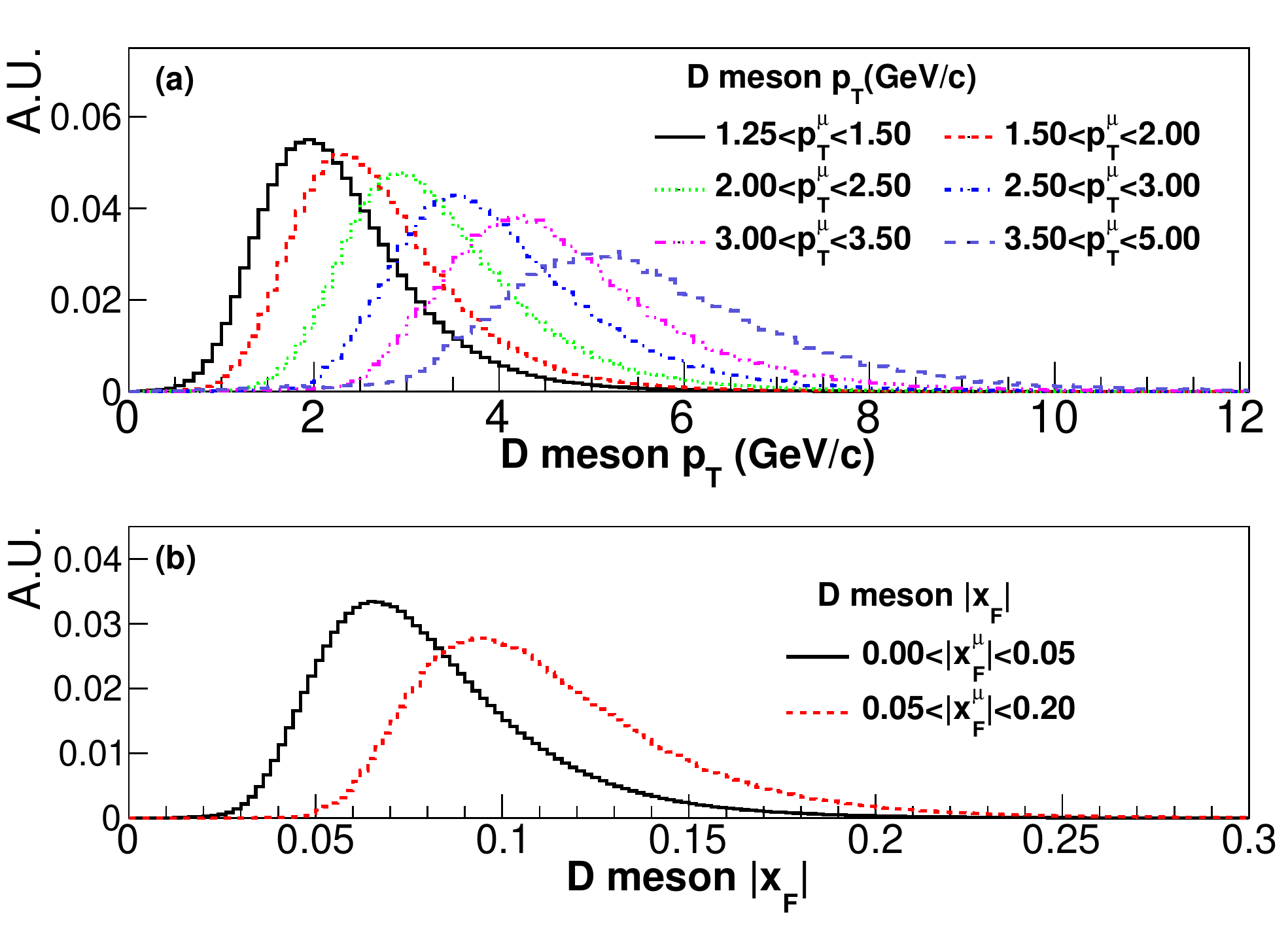}
\caption{\label{fig:charm_ptxf}
(a) \pt and (b) $|x_F|$ distributions of $D$ mesons ($D^0$, $\bar{D^0}$, 
$D^+$, and $D^-$) decaying into $\mu^{\pm}$ in $1.25<p_T^{\mu}<5.0$, 
$0.0<|x_F^{\mu}|<0.2$ and $1.4<|y^{\mu}|<2.0$ from {\sc pythia}. Each 
distribution is normalized to unity.}
\end{figure}

Figure~\ref{fig:charm_ptxf} shows the \pt and $|x_F|$ distributions of 
$D$ mesons which decay into muons in the kinematic range of this 
measurement ($1.25<p_T^{\mu}<5.0~{\rm GeV}/c$, $0.0<|x_F^{\mu}|<0.2$, 
and $1.4<|y^{\mu}|<2.0$); accepted charm hadrons comprise 
$D^0$(18.7\%), $\bar{D^0}$(20.3\%), $D^+$(24.2\%), $D^-$(26.1\%), and 
others ($D_s^+$, $D_s^-$, and baryons).  Because $A_N^{D^0}$ and 
$A_N^{D^+}$ ($A_N^{\bar{D^0}}$ and $A_N^{D^-}$) are very close in both 
models, the effect of potential different abundance of $D$ mesons 
between the data and {\sc pythia} is negligible. In addition, the 
modification of $A_N$ due to azimuthal smearing from the $D$-decay is 
quite small ($<5\%$ relative difference between $A_N^{D}$ and 
$A_N^{\mu}$) in $p_T^{\mu}>1.25~{\rm GeV}/c$. One notes that muons from 
charm and bottom are combined in the data, and the contribution from 
bottom is about 2\% (55\%) at $\pt=1~{\rm GeV}/c$ ($5~{\rm GeV}/c$) 
according to the FONLL calculation shown in 
Fig.~\ref{fig:HFmu_combined}. Therefore, the charm contribution is 
expected to be dominant except for the last \pt bin of $A_N^{\mu}$ 
($3.5<\pt<5~{\rm GeV}/c$). In addition, subprocesses other than 
gluon-fusion can contribute to the measured yield of muons from 
heavy-flavor decays. The converted $A_N$ of muons from $D$ mesons are 
shown in Fig.~\ref{fig:pt_AN_minus},~\ref{fig:pt_AN_plus}, 
and~\ref{fig:xf_AN_bothchg}, and both calculations are in agreement with 
the data within the statistical uncertainties. The difference between 
two models becomes larger at increasing $|x_F|$, but it is hard to 
distinguish these two models due to the limited $x_F$ coverage for this 
measurement ($\langle|x_F^{\mu}|\rangle$=0.04, 0.07).

\section{Summary}

We have reported the cross section and transverse single-spin asymmetry 
of muons from open heavy-flavor decays at $1.4<|y|<2.0$ in 
transversely-polarized \pp collisions at $\sqrts=200~{\rm GeV}$. 
Comparing with previous measurements by PHENIX, the cross section and 
asymmetry for positively-charged muons from open heavy-flavor decays are 
measured for the first time with the help of additional absorber 
material in the PHENIX muon arms. In the comparison with the FONLL 
calculation, the FONLL prediction is smaller than the measured cross 
section at low \pt where both experimental and theoretical systematic 
uncertainties are large, but it shows an agreement at 
$p_T>4~{\rm GeV}/c$ within systematic uncertainties.

Following the cross section results, we have measured the single-spin 
asymmetry of muons from open heavy-flavor decays for the first time. 
There is no clear indication of a nonzero asymmetry in the results, 
which have relatively large statistical uncertainties. Theoretical 
calculations of $A_N$ for $D$-meson production which take into account 
trigluon correlations are converted into $A_N$ for muons with the help 
of {\sc pythia} to compare directly with the data.  The 
calculations are in agreement with the data within experimental 
uncertainties.  Future studies with improved statistics (6.5 times 
current integrated luminosity of this analysis), using data taken with 
the PHENIX detector at RHIC in 2015, could provide further constraints 
on the trigluon correlation functions.


\section*{Acknowledgements}


We thank the staff of the Collider-Accelerator and Physics
Departments at Brookhaven National Laboratory and the staff of
the other PHENIX participating institutions for their vital
contributions.  
We also thank S. Yoshida and Y. Koike for the theory calculation.
We acknowledge support from the 
Office of Nuclear Physics in the
Office of Science of the Department of Energy,
the National Science Foundation, 
Abilene Christian University Research Council, 
Research Foundation of SUNY, and
Dean of the College of Arts and Sciences, Vanderbilt University 
(U.S.A),
Ministry of Education, Culture, Sports, Science, and Technology
and the Japan Society for the Promotion of Science (Japan),
Conselho Nacional de Desenvolvimento Cient\'{\i}fico e
Tecnol{\'o}gico and Funda\c c{\~a}o de Amparo {\`a} Pesquisa do
Estado de S{\~a}o Paulo (Brazil),
Natural Science Foundation of China (People's Republic of China),
Croatian Science Foundation and
Ministry of Science and Education (Croatia),
Ministry of Education, Youth, and Sports (Czech Republic),
Centre National de la Recherche Scientifique, Commissariat
{\`a} l'{\'E}nergie Atomique, and Institut National de Physique
Nucl{\'e}aire et de Physique des Particules (France),
Bundesministerium f\"ur Bildung und Forschung, Deutscher
Akademischer Austausch Dienst, and Alexander von Humboldt Stiftung (Germany),
National Science Fund, OTKA, EFOP, and the Ch. Simonyi Fund (Hungary),
Department of Atomic Energy and Department of Science and Technology (India), 
Israel Science Foundation (Israel), 
Basic Science Research Program through NRF of the Ministry of Education (Korea),
Physics Department, Lahore University of Management Sciences (Pakistan),
Ministry of Education and Science, Russian Academy of Sciences,
Federal Agency of Atomic Energy (Russia),
VR and Wallenberg Foundation (Sweden), 
the U.S. Civilian Research and Development Foundation for the
Independent States of the Former Soviet Union, 
the Hungarian American Enterprise Scholarship Fund,
and the US-Israel Binational Science Foundation.


\section*{APPENDIX:  DATA TABLES}

\begin{table*}[htb]
\caption{\label{tab:Xection}
Data table for the invariant cross section of muons from open heavy-flavor 
decays in $1.4<|y|<2.0$.}
\begin{ruledtabular}
\begin{tabular}{ccccccccc}
\pt (GeV/$c$) & $E\frac{d^{2}\sigma}{dp^{3}}$ (mb GeV$^{-2})$ & stat uncert. & syst uncert. &&
\pt (GeV/$c$) & $E\frac{d^{2}\sigma}{dp^{3}}$ (mb GeV$^{-2})$ & stat uncert. & syst uncert. \\\hline
1.375 & $7.9\times10^{-5}$ & $9.4\times10^{-7}$ & $2.4\times10^{-5}$ &&
3.25 & $3.1\times10^{-7}$ & $1.1\times10^{-8}$ & $4.5\times10^{-8}$\\
1.625 & $3.3\times10^{-5}$ & $3.7\times10^{-7}$ & $8.2\times10^{-6}$ &&
3.75 & $9.8\times10^{-8}$ & $5.0\times10^{-9}$ & $1.4\times10^{-8}$\\
1.875 & $1.2\times10^{-5}$ & $1.8\times10^{-7}$ & $2.9\times10^{-6}$ &&
4.25 & $3.2\times10^{-8}$ & $2.8\times10^{-9}$ & $4.7\times10^{-9}$\\
2.125 & $5.2\times10^{-6}$ & $1.0\times10^{-7}$ & $1.2\times10^{-6}$ &&
4.75 & $1.7\times10^{-8}$ & $1.8\times10^{-9}$ & $2.4\times10^{-9}$\\
2.375 & $2.4\times10^{-6}$ & $5.9\times10^{-8}$ & $4.7\times10^{-7}$ &&
5.5 & $4.5\times10^{-9}$ & $6.1\times10^{-10}$ & $6.5\times10^{-10}$\\
2.625 & $1.4\times10^{-6}$ & $3.8\times10^{-8}$ & $2.4\times10^{-7}$ &&
6.5 & $1.1\times10^{-9}$ & $3.3\times10^{-10}$ & $2.0\times10^{-10}$\\
2.875 & $6.8\times10^{-7}$ & $2.6\times10^{-8}$ & $1.1\times10^{-7}$ &&
& & & 
\end{tabular}
\end{ruledtabular}


\caption{\label{tab:AN_PT}
Data table for $A_{N}$ of muons from open heavy-flavor decays as a function of $p_T$.}
\begin{ruledtabular}
\begin{tabular}{ccccccccccc}
&& \multicolumn{3}{c}{Forward ($x_F>0$)} &&& \multicolumn{3}{c}{Backward ($x_F<0$)} \\
muon & $p_T$ bin (GeV/$c$) & $A_{N}$ & $\delta A_N^{\rm stat}$&  $\delta A_N^{\rm syst}$ &&
  $p_T$ bin (GeV/$c$) & $A_{N}$ & $\delta A_N^{\rm stat}$&  $\delta A_N^{\rm syst}$  \\ \hline
$\mu^{-}$ 
& (1.25, 1.50) & -0.101 & $\pm0.088$ & $^{+0.047}_{-0.095}$ &&
  (1.25, 1.50) & -0.138 & $\pm0.086$ & $^{+0.061}_{-0.146}$ \\
& (1.50, 2.00) & -0.003 & $\pm0.060$ & $^{+0.027}_{-0.027}$ &&
  (1.50, 2.00) & 0.110 & $\pm0.060$ & $^{+0.084}_{-0.047}$ \\
& (2.00, 2.50) & 0.045 & $\pm0.077$ & $^{+0.034}_{-0.027}$ &&
  (2.00, 2.50) & -0.060 & $\pm0.076$ & $^{+0.034}_{-0.051}$ \\
& (2.50, 3.00) & 0.016 & $\pm0.077$ & $^{+0.017}_{-0.016}$ &&
  (2.50, 3.00) & 0.022 & $\pm0.076$ & $^{+0.020}_{-0.019}$ \\
& (3.00, 3.50) & -0.056 & $\pm0.094$ & $^{+0.014}_{-0.015}$ &&
  (3.00, 3.50) & -0.002 & $\pm0.093$ & $^{+0.014}_{-0.014}$ \\
& (3.50, 5.00) & 0.087 & $\pm0.104$ & $^{+0.028}_{-0.025}$ && 
  (3.50, 5.00) & 0.018 & $\pm0.104$ & $^{+0.013}_{-0.013}$ \\ \\ 
$\mu^{+}$ 
& (1.25, 1.50) & 0.030 & $\pm0.069$ & $^{+0.035}_{-0.035}$ &&
  (1.25, 1.50) & -0.004 & $\pm0.066$ & $^{+0.033}_{-0.033}$ \\
& (1.50, 2.00) & -0.009 & $\pm0.040$ & $^{+0.026}_{-0.026}$ &&
  (1.50, 2.00) & -0.010 & $\pm0.039$ & $^{+0.025}_{-0.025}$ \\
& (2.00, 2.50) & 0.072 & $\pm0.055$ & $^{+0.036}_{-0.027}$ &&
  (2.00, 2.50) & -0.021 & $\pm0.054$ & $^{+0.025}_{-0.027}$ \\
& (2.50, 3.00) & 0.056 & $\pm0.065$ & $^{+0.028}_{-0.022}$ &&
  (2.50, 3.00) & -0.127 & $\pm0.066$ & $^{+0.034}_{-0.049}$ \\
& (3.00, 3.50) & 0.147 & $\pm0.087$ & $^{+0.038}_{-0.029}$ &&
  (3.00, 3.50) & -0.139 & $\pm0.088$ & $^{+0.033}_{-0.045}$ \\
& (3.50, 5.00) & -0.104 & $\pm0.108$ & $^{+0.035}_{-0.046}$ &&
  (3.50, 5.00) & -0.054 & $\pm0.109$ & $^{+0.016}_{-0.016}$ \\
\end{tabular}
\end{ruledtabular}


\caption{\label{tab:AN_XF}
Data table for $A_{N}$ of muons from open heavy-flavor decays as a function of $x_F$.}
\begin{ruledtabular}
\begin{tabular}{cccccccccccccc}
muon & $x_F$ bin  & $<x_F>$ & $A_N$& $\delta A_N^{\rm stat}$&  $\delta A_N^{\rm syst}$ &&
muon & $x_F$ bin & $<x_F>$ & $A_{N}$ & $\delta A_N^{\rm stat}$& $\delta A_N^{\rm syst}$  \\\hline
$\mu^{-}$
& (-0.20, -0.05) & -0.07 & 0.003 & $\pm0.048$ & $^{+0.007}_{-0.013}$ &&
$\mu^{+}$  
&  (-0.20, -0.05) & -0.07 & -0.030 & $\pm0.035$ & $^{+0.009}_{-0.014}$ \\
&  (-0.05, 0.00) & -0.04 & -0.009 & $\pm0.061$ & $^{+0.006}_{-0.010}$ &&
&  (-0.05, 0.00) & -0.04 & -0.026 & $\pm0.043$ & $^{+0.009}_{-0.026}$ \\
&  (0.00, 0.05) & 0.04 & -0.030 & $\pm0.062$ & $^{+0.010}_{-0.015}$ &&
&  (0.00, 0.05) & 0.04 & -0.004 & $\pm0.045$ & $^{+0.005}_{-0.013}$ \\
&  (0.05, 0.20) & 0.07 & 0.019 & $\pm0.047$ & $^{+0.009}_{-0.007}$ &&
&  (0.05, 0.20) & 0.07 & 0.058 & $\pm0.035$ & $^{+0.023}_{-0.013}$ \\
\end{tabular}
\end{ruledtabular}


\caption{\label{tab:AN_PT_SYS}
Sources of $\delta A_{N}^{\rm syst}$ for muons as a function of $p_T$.}
\begin{ruledtabular}
\begin{tabular}{ccccccccccccc}
&& \multicolumn{4}{c}{Forward ($x_F>0$)} &&& \multicolumn{4}{c}{Backward ($x_F<0$)} \\
muon  & $p_T$ bin (GeV/$c$) & $\delta{A_N^{\delta f_{\rm h}}}$ & $\delta{A_N^{{\rm h}}}$ &  $\delta{A_N^{J/\psi\to\mu}}$ & $\delta{A_N^{\rm method}}$  &&
$p_T$ bin (GeV/$c$) & $\delta{A_N^{\delta f_{\rm h}}}$ & $\delta{A_N^{{\rm h}}}$ &  $\delta{A_N^{J/\psi\to\mu}}$ & $\delta{A_N^{\rm method}}$  \\\hline
$\mu^{-}$ 
& (1.25, 1.50) & $^{+0.036}_{-0.090}$ & $^{+0.030}_{-0.030}$ & $^{+0.001}_{-0.000}$ & $^{+0.008}_{-0.008}$ &&
  (1.25, 1.50) & $^{+0.054}_{-0.143}$ & $^{+0.030}_{-0.030}$ & $^{+0.000}_{-0.000}$ & $^{+0.003}_{-0.003}$\\
& (1.50, 2.00) & $^{+0.003}_{-0.001}$ & $^{+0.026}_{-0.026}$ & $^{+0.001}_{-0.001}$ & $^{+0.004}_{-0.004}$ &&
  (1.50, 2.00) & $^{+0.079}_{-0.038}$ & $^{+0.027}_{-0.027}$ & $^{+0.001}_{-0.001}$ & $^{+0.007}_{-0.007}$\\
& (2.00, 2.50) & $^{+0.024}_{-0.012}$ & $^{+0.023}_{-0.023}$ & $^{+0.003}_{-0.003}$ & $^{+0.006}_{-0.006}$ &&
  (2.00, 2.50) & $^{+0.022}_{-0.044}$ & $^{+0.023}_{-0.023}$ & $^{+0.003}_{-0.003}$ & $^{+0.010}_{-0.010}$\\
& (2.50, 3.00) & $^{+0.004}_{-0.004}$ & $^{+0.014}_{-0.014}$ & $^{+0.005}_{-0.003}$ & $^{+0.007}_{-0.007}$ &&
  (2.50, 3.00) & $^{+0.009}_{-0.006}$ & $^{+0.014}_{-0.014}$ & $^{+0.004}_{-0.004}$ & $^{+0.010}_{-0.010}$\\
& (3.00, 3.50) & $^{+0.008}_{-0.011}$ & $^{+0.010}_{-0.010}$ & $^{+0.005}_{-0.004}$ & $^{+0.001}_{-0.001}$ &&
  (3.00, 3.50) & $^{+0.003}_{-0.004}$ & $^{+0.010}_{-0.010}$ & $^{+0.005}_{-0.005}$ & $^{+0.008}_{-0.008}$\\
& (3.50, 5.00) & $^{+0.018}_{-0.014}$ & $^{+0.009}_{-0.009}$ & $^{+0.007}_{-0.005}$ & $^{+0.019}_{-0.019}$ &&
  (3.50, 5.00) & $^{+0.001}_{-0.002}$ & $^{+0.009}_{-0.009}$ & $^{+0.006}_{-0.006}$ & $^{+0.007}_{-0.007}$\\
\\
$\mu^{+}$ 
& (1.25, 1.50) & $^{+0.007}_{-0.008}$ & $^{+0.034}_{-0.034}$ & $^{+0.000}_{-0.000}$ & $^{+0.007}_{-0.007}$ &&
  (1.25, 1.50) & $^{+0.001}_{-0.001}$ & $^{+0.032}_{-0.032}$ & $^{+0.000}_{-0.000}$ & $^{+0.001}_{-0.001}$\\
& (1.50, 2.00) & $^{+0.004}_{-0.007}$ & $^{+0.025}_{-0.025}$ & $^{+0.001}_{-0.001}$ & $^{+0.001}_{-0.001}$ &&
  (1.50, 2.00) & $^{+0.001}_{-0.003}$ & $^{+0.025}_{-0.025}$ & $^{+0.001}_{-0.001}$ & $^{+0.003}_{-0.003}$\\
& (2.00, 2.50) & $^{+0.028}_{-0.015}$ & $^{+0.023}_{-0.023}$ & $^{+0.003}_{-0.002}$ & $^{+0.003}_{-0.003}$ &&
  (2.00, 2.50) & $^{+0.005}_{-0.011}$ & $^{+0.022}_{-0.022}$ & $^{+0.002}_{-0.002}$ & $^{+0.011}_{-0.011}$\\
& (2.50, 3.00) & $^{+0.021}_{-0.014}$ & $^{+0.017}_{-0.017}$ & $^{+0.004}_{-0.003}$ & $^{+0.006}_{-0.006}$ &&
  (2.50, 3.00) & $^{+0.029}_{-0.046}$ & $^{+0.017}_{-0.017}$ & $^{+0.003}_{-0.003}$ & $^{+0.006}_{-0.006}$\\
& (3.00, 3.50) & $^{+0.035}_{-0.025}$ & $^{+0.013}_{-0.013}$ & $^{+0.005}_{-0.003}$ & $^{+0.007}_{-0.007}$ &&
  (3.00, 3.50) & $^{+0.027}_{-0.041}$ & $^{+0.013}_{-0.013}$ & $^{+0.004}_{-0.004}$ & $^{+0.012}_{-0.012}$\\
& (3.50, 5.00) & $^{+0.031}_{-0.043}$ & $^{+0.013}_{-0.013}$ & $^{+0.006}_{-0.004}$ & $^{+0.008}_{-0.008}$ &&
  (3.50, 5.00) & $^{+0.004}_{-0.004}$ & $^{+0.013}_{-0.013}$ & $^{+0.005}_{-0.005}$ & $^{+0.005}_{-0.005}$\\
\end{tabular}
\end{ruledtabular}
\end{table*}

\clearpage

\begin{table*}[ht!]
\caption{\label{tab:AN_XF_SYS}
Sources of $\delta A_{N}^{\rm syst}$ for muons as a function of $x_F$.}
\begin{ruledtabular}
\begin{tabular}{ccccccccccc}
muon & $x_F$ bin  & $\delta{A_N^{\delta f_{\rm h}}}$ & $\delta{A_N^{J/\psi\to\mu}}$ & $\delta{A_N^{\rm method}}$ &&
muon & $x_F$ bin  & $\delta{A_N^{\delta f_{\rm h}}}$ & $\delta{A_N^{J/\psi\to\mu}}$ & $\delta{A_N^{\rm method}}$ \\\hline
$\mu^{-}$ & (-0.20, -0.05) & $^{+0.003}_{-0.012}$ & $^{+0.005}_{-0.005}$ & $^{+0.003}_{-0.003}$ &&
$\mu^{+}$ & (-0.20, -0.05) & $^{+0.006}_{-0.013}$ & $^{+0.004}_{-0.004}$ & $^{+0.006}_{-0.006}$\\
&(-0.05, 0.00) & $^{+0.003}_{-0.008}$ & $^{+0.001}_{-0.001}$ & $^{+0.005}_{-0.005}$ &&
&(-0.05, 0.00) & $^{+0.009}_{-0.026}$ & $^{+0.001}_{-0.001}$ & $^{+0.002}_{-0.002}$\\
&(0.00, 0.05) & $^{+0.008}_{-0.013}$ & $^{+0.001}_{-0.001}$ & $^{+0.007}_{-0.007}$ &&
&(0.00, 0.05) & $^{+0.004}_{-0.013}$ & $^{+0.001}_{-0.001}$ & $^{+0.003}_{-0.003}$\\
&(0.05, 0.20) & $^{+0.005}_{-0.004}$ & $^{+0.005}_{-0.004}$ & $^{+0.005}_{-0.005}$ &&
&(0.05, 0.20) & $^{+0.022}_{-0.012}$ & $^{+0.004}_{-0.003}$ & $^{+0.005}_{-0.005}$\\
\end{tabular}
\end{ruledtabular}
\end{table*}



%
 
\end{document}